\documentclass[11pt,a4paper]{article}
\pdfoutput=1
\usepackage{jheppub}

\usepackage{color}
\usepackage{amsmath}
\usepackage{pifont}   
\usepackage{verbatim}       
\usepackage{acronym} 
    
\usepackage{amsfonts}
\usepackage{amssymb}
\usepackage{mathrsfs} 
\usepackage{graphicx}
\usepackage{multirow} 
\usepackage{slashed} 
\usepackage{array}

\usepackage{graphicx}
\usepackage{epsfig}
\usepackage{lscape}
\usepackage{rotating}
\usepackage{epsf}
\usepackage{bm}
\usepackage{booktabs}
\usepackage{array}
\usepackage{caption}
\usepackage{subcaption}
\usepackage[utf8]{inputenc}
\usepackage{float}

\usepackage{multirow}
\usepackage{array,booktabs,colortbl,multirow}
\usepackage{colortbl,xcolor,colordvi,color}
\usepackage{rotating}
\allowdisplaybreaks

\title{Contributions of inert electroweak multiplets to Higgs properties}

\author[a]{Hugues Beauchesne}
\author[b,a]{and Cheng-Wei Chiang}

\affiliation[a]{Physics Division, National Center for Theoretical Sciences,\\ Taipei 10617, Taiwan}
\affiliation[b]{Department of Physics and Center for Theoretical Physics, National Taiwan University, \\ Taipei 10617, Taiwan}

\emailAdd{beauchesneh@phys.ncts.ntu.edu.tw, chengwei@phys.ntu.edu.tw}

\abstract{New physics could manifest itself in the form of electroweak multiplets that interact at tree level with the Higgs boson but do not mix with Standard Model fields or acquire expectation values. In this paper, we study the potential contributions of such inert multiplets to several crucial Higgs properties, namely, the branching ratio of the Higgs to a $Z$ boson and a photon (or massless dark photon) and the triple Higgs coupling. Constraints from the Higgs signal strengths, oblique parameters and unitarity are taken into account.}

\begin{document}

\maketitle

\section{Introduction}\label{Sec:Intro}
Electroweak multiplets that neither mix with Standard Model (SM) fields nor acquire vacuum expectation values (VEVs) are ubiquitous in physics beyond the Standard Model (BSM). Such multiplets, which we will refer to as inert, can provide potential dark matter candidates \cite{LopezHonorez:2006gr, Gustafsson:2007pc, Huang:2017rzf, Betancur:2020fdl}, be responsible for baryogenesis \cite{Hambye:2007vf, Gil:2012ya}, or explain certain flavour anomalies~\cite{Chen:2023eof}.

Inert multiplets can in principle interact at tree level with the Higgs boson $h$ and therefore affect some of its properties. Interestingly, specifying these interaction terms and the masses of the particles involved is sufficient to compute their leading contributions to many Higgs properties. Since these interaction terms can only take a limited number of forms, this allows for a very broad analysis of the impact of inert multiplets on specific Higgs properties that are sensitive to new physics contributions. In this paper, we will study three such properties, all affected by the new particles at the one-loop level.

The first one is the branching ratio of the Higgs boson to a photon and a $Z$ boson. Since this is a loop process in the SM, new particles could potentially lead to sizable deviations from the SM prediction. Furthermore, the current measurement of this branching ratio is $2.1 \pm 0.7$ times larger than its SM value, corresponding to a deviation of $1.9\sigma$ \cite{ATLAS-CONF-2023-025, CMS-PAS-HIG-23-002}. Previous works on possible deviations of this branching ratio include Refs.~\cite{Archer-Smith:2020gib, Benbrik:2022bol}.

The second Higgs property of interest is the branching ratio of the Higgs to a $Z$ boson and a massless dark photon $A'$ \cite{Holdom:1985ag}. An extensive literature on the decay $h \to A A'$ already exists \cite{Gabrielli:2014oya, Biswas:2015sha, Biswas:2016jsh, Biswas:2017lyg, Biswas:2017anm, Beauchesne:2022svl, Beauchesne:2022fet, Beauchesne:2023bcy, Biswas:2022tcw}, but it was recently demonstrated in Refs.~\cite{Beauchesne:2022fet, Beauchesne:2023bcy, Biswas:2022tcw} that this branching ratio is considerably more constrained than the previous analysis Ref.~\cite{Gabrielli:2014oya}. Some of the constraints on $h \to A A'$ simply do not apply to $h \to Z A'$ and no bound on this branching ratio currently exists in the literature.

The third Higgs property considered is the triple Higgs coupling. This coupling provides additional information on the form of the Higgs potential. In addition, a larger value of this coupling could lead to a sufficiently strong first-order electroweak phase transition in the early Universe and thus explain matter abundance via electroweak baryogenesis~\cite{Kuzmin:1985mm, Cohen:1990it}.

More precisely, the goal of this paper is to study the potential contributions of inert multiplets to the Higgs branching ratios $\text{BR}(h \to Z A^{(\prime)})$ and the triple Higgs coupling. To do so, we will consider all possible tree-level interactions between the Higgs and inert multiplets and then compute their contributions to these Higgs properties. Constraints from the Higgs signal strengths, the electroweak precision tests and perturbative unitarity will be taken into account. In many ways, this paper is an extension of the formalism of Refs.~\cite{Beauchesne:2022fet, Beauchesne:2023bcy} and its application to new Higgs properties.

We find the following results. The branching ratio $\text{BR}(h \to Z A)$ can easily be enhanced by $\mathcal{O}(20\%)$ for simple models. Considerably larger enhancements are possible, but require more complicated models and careful fine-tuning. The branching ratio $\text{BR}(h \to Z A^\prime)$ could in principle be above $1\%$. It would however require new neutral particles barely above half the mass of the Higgs boson. A branching ratio of $\mathcal{O}(0.1\%)$ is however relatively easy to obtain. The triple Higgs coupling can easily be enhanced by a factor of several.

This paper is organized as follows. We introduce in Sec.~\ref{Sec:Vertices} the inert multiplets and all their relevant interactions with the Higgs doublet. Sec.~\ref{Sec:Constraints and observables} explains how we apply constraints and compute the above-mentioned properties. The results are shown in Sec.~\ref{Sec:Results}. Some concluding remarks are presented in Sec.~\ref{Sec:Conclusion}.

\section{Possible interaction terms}\label{Sec:Vertices}
If inert multiplets are to have any sizable impact on Higgs properties, they need to interact with the Higgs boson at tree level. In this section, we present all renormalizable interaction terms relevant to our observables. Most of these terms were already included in Refs.~\cite{Beauchesne:2022fet, Beauchesne:2023bcy}, from which we borrow heavily. Contrary to these papers, we will not in general require the inert multiplets to be charged under a new $U(1)'$ symmetry. This will allow for a few additional terms as well as real multiplets. We will require the interaction terms to be able to contribute to our observables at one loop. In combination with the requirement of the fields not acquiring VEVs, this will force the interaction terms to always contain exactly two (distinct or not) inert multiplets. We will write quantum numbers of fields as $(SU(2)_L, U(1)_Y, U(1)')$. The $SU(2)_L$ indices are labeled by Latin letters and range from 1 to the size of the representation of the corresponding field. Considering the constraints on new coloured particles, all fields will be assumed to be $SU(3)_C$ singlets. Terms that are equivalent to those below up to a field redefinition are not included in the list, but could easily be included when the field redefinition cannot be performed simultaneously for all terms in the Lagrangian. See Ref.~\cite{Banta:2021dek} for a similar exercise but with different assumptions and motivations.

\noindent{\it Fermion case}: Consider the fermions
\begin{equation}\label{eq:FieldsFermionsI}
  \psi_1: (\mathbf{p}, Y^p, Q'), \qquad \psi_2: (\mathbf{n}, Y^n, Q'),
\end{equation}
with $Y^p = Y^n + 1/2$ and $p = n \pm 1$.  Throughout the paper, we use the boldfaced $\bf p$ and $\bf n$ to denote the $SU(2)_L$ representation of a multiplet and italic $p$ and $n$ for their corresponding dimensions. The following term is allowed\footnote{Summation over repeated symbols is implicit when trivial but will be written explicitly otherwise.}
\begin{equation}\label{eq:LagrangianFI}
  F(\psi_1, \psi_2) = -\hat{d}^{pn}_{abc} \overline{\psi}_1^a (A_L P_L + A_R P_R) \psi_2^b H^c + \text{H.c.}
\end{equation}
The $SU(2)_L$ tensor is given by the Clebsch-Gordan coefficient
\begin{equation}\label{eq:SU2TensorF1a}
  \hat{d}^{pn}_{abc} = C^{JM}_{j_1 m_1 j_2 m_2} \equiv \langle j_1 j_2 m_1 m_2 |J M \rangle,
\end{equation}
where
\begin{equation}\label{eq:SU2TensorF1b}
  \begin{aligned}
    j_1 &= \frac{n - 1}{2},     & j_2 &= \frac{1}{2},     & J &= \frac{p -1}{2},       \\ 
    m_1 &= \frac{n + 1 - 2b}{2}, & m_2 &=\frac{3 - 2c}{2}, & M &= \frac{p + 1 - 2a}{2}.
  \end{aligned}
\end{equation}
Depending on their quantum numbers, either $\psi_1$ or $\psi_2$ can be a real multiplet.

\noindent{\it Scalar case I}: Consider the scalars
\begin{equation}\label{eq:FieldsScalarsI}
  \phi_1: (\mathbf{p}, Y^p, Q'), \qquad \phi_2: (\mathbf{n}, Y^n, Q'),
\end{equation}
with $Y^p = Y^n + 1/2$ and $p = n \pm 1$. The following term is allowed
\begin{equation}\label{eq:LagrangianSI}
  S_1(\phi_1, \phi_2) = -\mu \hat{d}_{abc}^{pn} \phi_1^{a\dagger}\phi_2^b H^c + \text{H.c.}
\end{equation}
The $SU(2)_L$ tensor is given by
\begin{equation}\label{eq:SU2TensorSIa}
  \hat{d}^{pn}_{abc} = C^{JM}_{j_1 m_1 j_2 m_2},
\end{equation}
where
\begin{equation}\label{eq:SU2TensorSIb}
  \begin{aligned}
    j_1 &= \frac{n - 1}{2},     & j_2 &= \frac{1}{2},     & J &= \frac{p -1}{2},       \\ 
    m_1 &= \frac{n + 1 - 2b}{2}, & m_2 &=\frac{3 - 2c}{2}, & M &= \frac{p + 1 - 2a}{2}.
  \end{aligned}
\end{equation}
Depending on their quantum numbers, either $\phi_1$ or $\phi_2$ can be a real multiplet. This case includes the electroweak (EW) case of Refs.~\cite{Gabrielli:2014oya, Biswas:2015sha, Biswas:2016jsh, Biswas:2017lyg, Biswas:2017anm} when their neutral scalar $S$ is replaced by its expectation value.

\noindent{\it Scalar case II}: Consider the scalar
\begin{equation}\label{eq:FieldsScalarsII}
  \phi: (\mathbf{n}, Y^n, Q').
\end{equation}
The following term is allowed
\begin{equation}\label{eq:LagrangianSII}
  S_2(\phi) = -\sum_{r \in \{n - 1, n + 1\}}\lambda^r \hat{d}_{abcd}^{nr} H^{a\dagger} H^b \phi^{c\dagger}\phi^d.
\end{equation}
The $SU(2)_L$ tensor is given by
\begin{equation}\label{eq:SU2TensorSIIa}
  \hat{d}^{nr}_{abcd} = \sum_M C^{JM}_{j_1 m_1 j_2 m_2} C^{JM}_{j_3 m_3 j_4 m_4},
\end{equation}
where $M$ is summed over $\{-J, -J +1, -J + 2, ..., +J\}$ and
\begin{equation}\label{eq:SU2TensorSIIb}
  \begin{aligned}
    j_1 &= \frac{1}{2},      & j_2 &= \frac{n - 1}{2},      & j_3 &= \frac{1}{2},      & j_4 &= \frac{n - 1}{2}, & J &= \frac{r - 1}{2},\\
    m_1 &= \frac{3 - 2a}{2}, & m_2 &= \frac{n + 1 - 2c}{2}, & m_3 &= \frac{3 - 2b}{2}, & m_4 &= \frac{n + 1 - 2d}{2}.\\ 
  \end{aligned}
\end{equation}
Unless $\phi$ is a singlet, there are in general two possible contractions of the $SU(2)_L$ indices and therefore two coefficients. Depending on its quantum numbers, $\phi$ can be either a real or complex multiplet.

\noindent{\it Scalar case III}: Consider the scalars
\begin{equation}\label{eq:FieldsScalarsIII}
  \phi_1: (\mathbf{p}, Y^p, Q'), \qquad \phi_2: (\mathbf{n}, Y^n, Q'),
\end{equation}
with $p \in \{n - 2, n, n + 2\}$ and $Y^p = Y^n$. The following term is allowed
\begin{equation}\label{eq:LagrangianSIII}
  S_3(\phi_1, \phi_2) = -\sum_{r \in \mathcal{R}}\lambda^r \hat{d}_{abcd}^{pnr} H^{a\dagger} H^b \phi^{c\dagger}_1\phi^d_2 + \text{H.c.},
\end{equation}
where $\mathcal{R}=\{n - 1, n + 1\}\cap \{p - 1, p + 1\}$. The $SU(2)_L$ tensor is given by
\begin{equation}\label{eq:SU2TensorSIIIa}
  \hat{d}^{pnr}_{abcd} = \sum_M C^{JM}_{j_1 m_1 j_2 m_2} C^{JM}_{j_3 m_3 j_4 m_4},
\end{equation}
where $M$ is summed over $\{-J, -J + 1, -J + 2, ..., +J\}$ and
\begin{equation}\label{eq:SU2TensorSIIIb}
  \begin{aligned}
    j_1 &= \frac{1}{2},      & j_2 &= \frac{p - 1}{2},     & j_3 &= \frac{1}{2},      & j_4 &= \frac{n - 1}{2},      & J &= \frac{r - 1}{2},\\
    m_1 &= \frac{3 - 2a}{2}, & m_2 &= \frac{p + 1 - 2c}{2} & m_3 &= \frac{3 - 2b}{2}, &  m_4 &= \frac{n + 1 - 2d}{2}.\\
  \end{aligned}
\end{equation}
If $p$ and $n$ differ by two, there is only one possible contraction. If $p = n$, there are two possible contractions unless $p = n = 1$. Depending on their quantum numbers, either $\phi_1$ and $\phi_2$ can be a real multiplet.

\noindent{\it Scalar case IV}: Consider the scalars
\begin{equation}\label{eq:FieldsScalarsIV}
  \phi_1: (\mathbf{p}, Y^p, Q'), \qquad \phi_2: (\mathbf{n}, Y^n, Q'),
\end{equation}
with $p \in \{n - 2, n, n + 2\}$ and $Y^p = Y^n + 1$. The following term is allowed
\begin{equation}\label{eq:LagrangianSIV}
  S_4(\phi_1, \phi_2) = -\lambda \hat{d}_{abcd}^{pn} H^a H^b \phi^{c\dagger}_1\phi^d_2 + \text{H.c.}
\end{equation}
The $SU(2)_L$ tensor is given by
\begin{equation}\label{eq:SU2TensorSIVa}
  \hat{d}^{pn}_{abcd} = \sum_{M_1} C^{J_1 M_1}_{j_1 m_1 j_2 m_2} C^{J_2 M_2}_{J_1 M_1 j_3 m_3},
\end{equation}
where $M_1$ is summed over $\{-1, 0, 1\}$ and
\begin{equation}\label{eq:SU2TensorSIVb}
  \begin{aligned}
    j_1 &= \frac{1}{2},      & j_2 &= \frac{1}{2},     & j_3 &= \frac{n - 1}{2},      & J_1 = 1,\;\; & J_2 = \frac{p - 1}{2},\\
    m_1 &= \frac{3 - 2a}{2}, & m_2 &= \frac{3 - 2b}{2} & m_3 &= \frac{n + 1 - 2d}{2}, &              & M_2 = \frac{p + 1 -2c}{2}.\\
  \end{aligned}
\end{equation}
Only one contraction of the $SU(2)_L$ indices is allowed. Depending on their quantum numbers, either $\phi_1$ or $\phi_2$ can be a real multiplet.

\noindent{\it Scalar case V}: Consider the scalar
\begin{equation}\label{eq:FieldsScalarsV}
  \phi: (\mathbf{n}, 1/2, 0).
\end{equation}
The following term is allowed
\begin{equation}\label{eq:LagrangianSV}
  S_5(\phi) = -\lambda \hat{d}_{abcd}^n H^a H^b \phi^{c\dagger}\phi^{d\dagger} + \text{H.c.}
\end{equation}
The $SU(2)_L$ tensor is given by
\begin{equation}\label{eq:SU2TensorSVa}
  \hat{d}^n_{abcd} = \sum_M C^{J M}_{j_1 m_1 j_2 m_2} C^{J M}_{J_3 m_3 j_4 m_4},
\end{equation}
where $M$ is summed over $\{-1, 0, 1\}$ and
\begin{equation}\label{eq:SU2TensorSVb}
  \begin{aligned}
    j_1 &= \frac{1}{2},      & j_2 &= \frac{1}{2},     & j_3 &= \frac{n - 1}{2},      & j_4 &= \frac{n - 1}{2},      & J = 1, \\
    m_1 &= \frac{3 - 2a}{2}, & m_2 &= \frac{3 - 2b}{2} & m_3 &= \frac{n + 1 - 2c}{2}, & m_4 &= \frac{n + 1 - 2d}{2}. &        \\
  \end{aligned}
\end{equation}
Only one non-zero contraction of the $SU(2)_L$ indices is possible. The scalar $\phi$ must be a complex multiplet. It must also be of even dimension to be non-zero, which can easily be verified from a table of Clebsch-Gordan coefficients. This interaction term does not allow $\phi$ to be charged under $U(1)'$ and thus did not appear in Refs.~\cite{Beauchesne:2022fet, Beauchesne:2023bcy}.

\noindent{\it Scalar case VI}: Consider the scalar
\begin{equation}\label{eq:FieldsScalarsVI}
  \phi: (\mathbf{n}, 0, 0).
\end{equation}
The following term is allowed
\begin{equation}\label{eq:LagrangianSVI}
  S_6(\phi) = -\sum_{r \in \{1, 3\}}\lambda^r \hat{d}_{abcd}^{nr} H^{a\dagger} H^b \phi^c\phi^d + \text{H.c.}
\end{equation}
The $SU(2)_L$ tensor is given by
\begin{equation}\label{eq:SU2TensorSVIa}
  \hat{d}^{nr}_{abcd} = \sum_M C^{J_1 M}_{j_1 m_1 j_2 m_2} C^{J_2 M_2}_{J_1 M j_3 m_3},
\end{equation}
where $M$ is summed over $\{-J_1, -J_1 + 1, -J_1 + 2,..., +J_1\}$ and
\begin{equation}\label{eq:SU2TensorSVIb}
  \begin{aligned}
    j_1 &= \frac{n - 1}{2},      & j_2 &= \frac{n - 1}{2},     & j_3 &= \frac{1}{2},      & J_1 &= \frac{r - 1}{2}, & J_2 =& \frac{1}{2},\\
    m_1 &= \frac{n + 1 - 2c}{2}, & m_2 &= \frac{n + 1 - 2d}{2} & m_3 &= \frac{3 - 2b}{2}, & & & M_2 =& \frac{3 - 2a}{2}.        \\
  \end{aligned}
\end{equation}
Two contractions of the $SU(2)_L$ indices are possible, though in practice only $r=1$ contributes if $n$ is odd and only $r=3$ if $n$ is even. The scalar $\phi$ can be either real or complex. This term was not permitted by the requirements of Refs.~\cite{Beauchesne:2022fet, Beauchesne:2023bcy}.

\section{Constraints and observables}\label{Sec:Constraints and observables}
In this section, we introduce useful notation, explain how constraints are applied and present the computations of relevant Higgs properties. Many results from this section can be found in Refs.~\cite{Beauchesne:2022fet, Beauchesne:2023bcy}, though we modified the notation and expanded them to apply to additional scenarios.

\subsection{Lagrangian}\label{sSec:Lagrangian}
The only parts of the Lagrangian necessary to compute our observables are the kinematic terms and the interaction terms of the last section. In this section, we will write down the most general form of the relevant Lagrangian terms that can result from them. All new particles considered will be either complex scalars $\phi_i^C$, real scalars $\phi_i^R$, Dirac fermions $\psi_i^D$ or Majorana fermions $\psi_i^M$, the only exception being a potential dark photon $A'$. The Lagrangian is expressed in terms of gauge eigenstates. Many parameters will be introduced for convenience, though some are related either by construction or gauge symmetries.

\noindent{\it Mass terms}:
\begin{equation}\label{eq:LagrangianMass}
  \begin{aligned}
  \mathcal{L}^{\text{Mass}} = & -(m_C^2)_{ij} {\phi_i^C}^\dagger \phi_j^C - \frac{1}{2}(m_R^2)_{ij} \phi_i^R \phi_j^R\\
                              & - \bar{\psi}^D_i\left((m^L_D)_{ij} P_L +  (m^R_D)_{ij} P_R \right)\psi^D_j - \frac{1}{2}\bar{\psi}^M_i\left((m^L_M)_{ij} P_L +  (m^R_M)_{ij} P_R  \right)\psi^M_j.
  \end{aligned}
\end{equation}

\noindent{\it Yukawa interactions}:
\begin{equation}\label{eq:LagrangianYukawa}
  \begin{aligned}
    \mathcal{L}^{\text{Yukawa}} = & - h (\Omega_{C})_{ij} {\phi_i^C}^\dagger \phi_j^C - \frac{h}{2} (\Omega_{R})_{ij} \phi_i^R \phi_j^R\\
                                  & - h \bar{\psi}^D_i\left((\Omega^L_D)_{ij} P_L +  (\Omega^R_D)_{ij} P_R \right)\psi^D_j - \frac{h}{2}\bar{\psi}^M_i\left((\Omega^L_M)_{ij} P_L +  (\Omega^R_M)_{ij} P_R  \right)\psi^M_j.
  \end{aligned}
\end{equation}

\noindent{\it Gauge interactions of complex scalars}:
\begin{equation}\label{eq:LagrangianGaugeCS}
  \begin{aligned}
    \mathcal{L}^{\text{Gauge}}_C = \;\;\;&\left(-i (A_C)_{ii} A_\mu {\phi^C_i}^\dagger \partial^\mu \phi^C_i + \text{H.c.}\right) + (B_C)_{ii} A_\mu A^\mu {\phi^C_i}^\dagger \phi^C_i\\
    + &\left(-i (C_C)_{ij} Z_\mu {\phi^C_i}^\dagger \partial^\mu \phi^C_j + \text{H.c.}\right) + (D_C)_{ij} Z_\mu Z^\mu {\phi^C_i}^\dagger \phi^C_j
    + 2 (E_C)_{ij} A_\mu Z^\mu {\phi^C_i}^\dagger \phi^C_j\\
    + &\left(-i (F_C)_{ij}W^+_\mu\left({\phi^C_i}^\dagger \partial^\mu \phi^C_j - \partial^\mu{\phi^C_i}^\dagger \phi^C_j\right) + \text{H.c.}\right) 
    + 2 (G_C)_{ij}W^+_\mu {W^-}^\mu {\phi^C_i}^\dagger \phi^C_j\\
    + &\left(-i\frac{(H_C^1)_{ij}}{\sqrt{2}}W^+_\mu(\phi_i^{C\dagger}\partial^\mu\phi_j^{C\dagger} - \partial^\mu\phi_i^{C\dagger}\phi_j^{C\dagger}) +\text{H.c.}\right)\\
    + &\left(-i\frac{(H_C^2)_{ij}}{\sqrt{2}}W^+_\mu(\phi_i^C\partial^\mu\phi_j^C - \partial^\mu\phi_i^C\phi_j^C)+\text{H.c.}\right)\\
   + &\left(-i (A'_C)_{ii} A'_\mu {\phi^C_i}^\dagger \partial^\mu \phi^C_i + \text{H.c.}\right) + 2 (B'_C)_{ii} A'_\mu A^\mu {\phi^C}^\dagger_i \phi^C_i
   + (B''_C)_{ii} A'_\mu A'^\mu {\phi^C}^\dagger_i \phi^C_i\\
   + &\; 2 (E'_C)_{ij} A'_\mu Z^\mu {\phi^C_i}^\dagger \phi^C_j.
  \end{aligned}
\end{equation}

\noindent{\it Gauge interactions of real scalars}:
\begin{equation}\label{eq:LagrangianGaugeRS}
  \mathcal{L}^{\text{Gauge}}_R = \frac{(C_R)_{ij}}{2} Z_\mu\left(\phi^R_i \partial^\mu\phi^R_j - \partial^\mu\phi^R_i \phi^R_j\right) + \frac{(D_R)_{ij}}{2} Z_\mu Z^\mu \phi^R_i \phi^R_j + (G_R)_{ij}W^+_\mu {W^-}^\mu \phi^R_i \phi^R_j.
\end{equation}

\noindent{\it Gauge interactions of both complex and real scalars}:
\begin{equation}\label{eq:LagrangianGaugeRC}
  \mathcal{L}^{\text{Gauge}}_{RC} = -i (F^1_{RC})_{ij} W^+_\mu \left(\phi^R_i \partial^\mu \phi^C_j - \partial^\mu\phi^R_i \phi^C_j\right) 
                                               - i (F^2_{RC})_{ij} W^+_\mu \left({\phi^C_i}^\dagger \partial^\mu \phi^R_j - \partial^\mu{\phi^C_i}^\dagger \phi^R_j\right)+ \text{H.c.}
\end{equation}

\noindent{\it Gauge interactions of Dirac fermions}:
\begin{equation}\label{eq:LagrangianDirac}
  \begin{aligned}
    \mathcal{L}^{\text{Gauge}}_D = & - (A_D)_{ii} A_\mu \bar{\psi}^D_i \gamma^\mu \psi^D_i - Z_\mu \bar{\psi}^D_i\gamma^\mu\left((B^L_D)_{ij} P_L + (B^R_D)_{ij} P_R\right)\psi^D_j\\
                                          & - W^+_\mu \bar{\psi}^D_i\gamma^\mu\left((F^L_D)_{ij} P_L + (F^R_D)_{ij} P_R\right)\psi^D_j + \text{H.c.}\\
                                          & - (A'_D)_{ii} A'_\mu \bar{\psi}^D_i \gamma^\mu \psi^D_i.
  \end{aligned}
\end{equation}

\noindent{\it Gauge interactions of Majorana fermions}:
\begin{equation}\label{eq:LagrangianMajorana}
  \mathcal{L}^{\text{Gauge}}_M = - \frac{Z_\mu}{2} \bar{\psi}^M_i\gamma^\mu\left((F^L_M)_{ij} P_L + (F^R_M)_{ij} P_R\right)\psi^M_j.
\end{equation}

\noindent{\it Gauge interactions of both Dirac and Majorana fermions}:
\begin{equation}\label{eq:LagrangianDM}
  \begin{aligned}
    \mathcal{L}^{\text{Gauge}}_{MD} = &- W^+_\mu \bar{\psi}^M_i\gamma^\mu\left((F^{1L}_{MD})_{ij} P_L + (F^{1R}_{MD})_{ij} P_R\right)\psi^D_j + \text{H.c.} \\
                                       &- W^+_\mu \bar{\psi}^D_i\gamma^\mu\left((F^{2L}_{MD})_{ij} P_L + (F^{2R}_{MD})_{ij} P_R\right)\psi^M_j + \text{H.c.} \\
  \end{aligned}
\end{equation}

\noindent{\it Quartic couplings}:
\begin{equation}\label{eq:LagrangianQuartic}
  \mathcal{L}^{\text{Quartic}} = -\frac{\lambda^C_{ij}}{2} h^2 \phi^{C\dagger}_i\phi^C_j - \frac{\lambda^R_{ij}}{4} h^2 \phi^R_i\phi^R_j.
\end{equation}

The different powers of 2 that appear in these equations are chosen to simplify the results of the following sections. In practice, we compute analytically the different coefficients of the Lagrangian via a simple program. The mass matrices of Eq.~\eqref{eq:LagrangianMass} are then diagonalized numerically. The resulting mass eigenstates are labelled with a hat and their corresponding masses are referred to as $\hat{m}_i^\alpha$, with $\alpha$ being either $C$, $R$, $D$ or $M$ as in Eq.~\eqref{eq:LagrangianMass}. The coefficients in the mass eigenstates basis are also labeled with a hat. The $H^1_C$ and $H^2_C$ terms only appear in very exotic cases.

\subsection{Relevant Higgs decays and Higgs signal strengths}\label{sSec:HiggsDecay}
We present in this section all relevant decay widths of the Higgs boson. We also discuss our implementation of the constraints on the Higgs signal strengths.

\subsubsection{Higgs to (dark) photons}\label{ssSec:htoAA}
Only complex scalars and Dirac fermions can contribute at one loop to the Higgs decay to photons or dark photons. The amplitudes take the general form
\begin{equation}\label{eq:AmplitudehtoAA}
  M^{h\to A^{(\prime)}A^{(\prime)}} =  S^{h\to A^{(\prime)}A^{(\prime)}} \left(p_1\cdot p_2 g_{\mu\nu} - p_{1\mu} p_{2\nu}\right)\epsilon^\nu_{p_1}\epsilon^\mu_{p_2}
                                     + i\tilde{S}^{h\to A^{(\prime)}A^{(\prime)}}\epsilon_{\mu\nu\alpha\beta}p_1^\alpha p_2^\beta\epsilon^\nu_{p_1}\epsilon^\mu_{p_2},
\end{equation}
with $p_i$ and $\epsilon_i$ being respectively the momentum and polarization of the gauge bosons. The new physics contributions to the coefficients are
\begin{equation}\label{eq:SStildeAA}
  \begin{aligned}
     4\pi^2S^{h\to A^{(\prime)}A^{(\prime)}}         &= \sum_i (\hat{A}_C^{(\prime)})_{ii} (\hat{A}_C^{(\prime)})_{ii} (\hat{\Omega}_C)_{ii} f_1(\hat{m}^C_i) 
                                                      + \sum_i (\hat{A}_D^{(\prime)})_{ii} (\hat{A}_D^{(\prime)})_{ii} \text{Re}((\hat{\Omega}^L_D)_{ii}) f_2(\hat{m}_i^D),\\
     4\pi^2\tilde{S}^{h\to A^{(\prime)}A^{(\prime)}} &= \sum_i (\hat{A}_D^{(\prime)})_{ii} (\hat{A}_D^{(\prime)})_{ii} \text{Im}((\hat{\Omega}^L_D)_{ii}) f_3(\hat{m}_i^D),
  \end{aligned}
\end{equation}
where
\begin{equation}\label{eq:f1f2f3}
  \begin{aligned}
    f_1(m_i) &=  \frac{1 + 2m_i^2 C_0(0, 0, m_h^2; m_i, m_i, m_i)}{m_h^2},\\
    f_2(m_i) &= -2\frac{m_i}{m_h^2}\left[2 + (4 m_i^2 - m_h^2)C_0(0, 0, m_h^2; m_i, m_i, m_i) \right],\\
    f_3(m_i) &= -2i m_i C_0(0, 0, m_h^2; m_i, m_i, m_i),    
  \end{aligned}
\end{equation}
with $C_0(s_1, s_{12}, s_2; m_0, m_1, m_2)$ being the scalar three-point Passarino-Veltman function \cite{Passarino:1978jh}.\footnote{All loop computations were performed with the assistance of Package-X \cite{Patel:2015tea}.} In obtaining these results, we have used the fact that certain coefficients are related by gauge symmetries (e.g., $(\hat{B}_C)_{ii} = (\hat{A}_C)_{ii}^2$) and also $(\hat{\Omega}^R_D)_{ii}^\ast = (\hat{\Omega}^L_D)_{ii}$. The decay widths are then
\begin{equation}\label{eq:hAAwidth}
  \Gamma^{h \to A^{(\prime)}A^{(\prime)}} = \frac{|S^{h\to A^{(\prime)}A^{(\prime)}}|^2 + |\tilde{S}^{h\to A^{(\prime)}A^{(\prime)}}|^2}{32\pi n_S}m_h^3,
\end{equation}
where $n_S$ is 2 for $AA$ and $A'A'$ and 1 for $AA'$.

\subsubsection{Higgs decay to $Z$ and a (dark) photon}\label{ssSec:htoAZ}
Once again, only complex scalars and Dirac fermions can contribute to the Higgs decay to a $Z$ boson and a photon or dark photon. The amplitudes take the general form
\begin{equation}\label{eq:AmplitudehtoAZ}
  M^{h\to A^{(\prime)}Z} =  S^{h\to A^{(\prime)}Z} \left(p_1\cdot p_2 g_{\mu\nu} - p_{1\mu} p_{2\nu}\right)\epsilon^\nu_{p_1}\epsilon^\mu_{p_2}
                                     + i\tilde{S}^{h\to A^{(\prime)}Z}\epsilon_{\mu\nu\alpha\beta}p_1^\alpha p_2^\beta\epsilon^\nu_{p_1}\epsilon^\mu_{p_2}.
\end{equation}
The new physics contributions to the coefficients are
\begin{equation}\label{eq:SStildeAZ}
  \begin{aligned}
     4\pi^2 S^{h\to A^{(\prime)}Z}         =& \sum_{i, j} (\hat{A}_C^{(\prime)})_{ii} (C_C)_{ji} (\hat{\Omega}_C)_{ij} f_4(\hat{m}^C_i, \hat{m}^C_j)\\
                                           &+ \sum_{i, j} (\hat{A}^{(\prime)}_D)_{ii}\left[(\hat{B}^L_D)_{ji}(\hat{\Omega}^L_D)_{ij} + (\hat{B}^R_D)_{ji}(\hat{\Omega}^R_D)_{ij}\right]f_5(\hat{m}^D_i, \hat{m}^D_j)\\
                                           &+ \sum_{i, j} (\hat{A}^{(\prime)}_D)_{ii}\left[(\hat{B}^R_D)_{ji}(\hat{\Omega}^L_D)_{ij} + (\hat{B}^L_D)_{ji}(\hat{\Omega}^R_D)_{ij}\right]f_5(\hat{m}^D_j, \hat{m}^D_i),
\\
     4\pi^2 \tilde{S}^{h\to A^{(\prime)}Z} =& \sum_{i, j} (\hat{A}^{(\prime)}_D)_{ii}\left[(\hat{B}^L_D)_{ji}(\hat{\Omega}^L_D)_{ij} - (\hat{B}^R_D)_{ji}(\hat{\Omega}^R_D)_{ij}\right]f_6(\hat{m}^D_i, \hat{m}^D_j)\\
                                           &+ \sum_{i, j} (\hat{A}^{(\prime)}_D)_{ii}\left[(\hat{B}^R_D)_{ji}(\hat{\Omega}^L_D)_{ij} - (\hat{B}^L_D)_{ji}(\hat{\Omega}^R_D)_{ij}\right]f_6(\hat{m}^D_j, \hat{m}^D_i),
  \end{aligned}
\end{equation}
where
\begin{equation}\label{eq:f4f5f6}
  \begin{aligned}
    f_4(m_i, m_j) =& \frac{1}{m_h^2 - m_Z^2}\Biggl[1 + \frac{m_i^2 - m_j^2}{2m_h^2}\ln\frac{m_i^2}{m_j^2} + \frac{m_Z^2\left(\Lambda(m_h^2, m_i, m_j) - \Lambda(m_Z^2, m_i, m_j)\right)}{m_h^2 - m_Z^2}\\
                   & \hspace{2.0cm} + m_i^2 C_0(0, m_h^2, m_Z^2; m_i, m_i, m_j) + m_j^2 C_0(0, m_h^2, m_Z^2; m_j, m_j, m_i)\Biggr]\\
    f_5(m_i, m_j) =& -m_i f_4(m_i, m_j) + \frac{m_i}{2}C_0(0, m_h^2, m_Z^2; m_i, m_i, m_j)\\
    f_6(m_i, m_j) =& -\frac{m_i}{2} C_0(0, m_h^2, m_Z^2; m_i, m_i, m_j),
  \end{aligned}
\end{equation}
and $\Lambda(s; m_i, m_j)$ is defined as
\begin{equation}\label{eq:LambdaDef}
  \Lambda(s; m_i, m_j) = \lim_{\epsilon \to 0^+}\frac{\sqrt{\lambda(s, m_i^2, m_j^2)}}{s}\ln\left(\frac{m_i^2 + m_j^2 - s + \sqrt{\lambda(s, m_i^2, m_j^2)}}{2 m_i m_j} + i\epsilon\right),
\end{equation}
with $\lambda(x, y, z) = x^2 + y^2 + z^2 - 2xy -2xz - 2yz$ being the K\"all\'en function. The decay widths are then
\begin{equation}\label{eq:hAZwidth}
  \Gamma^{h \to A^{(\prime)}Z} = \frac{|S^{h\to A^{(\prime)}Z}|^2 + |\tilde{S}^{h\to A^{(\prime)}Z}|^2}{32\pi }\left(1 - \frac{m_Z^2}{m_h^2}\right)^3 m_h^3.
\end{equation}
Note that one major qualitative difference between these decay widths and $\Gamma^{h \to A^{(\prime)}A^{(\prime)}}$ is the suppression factor of $\left(1 - m_Z^2/m_h^2\right)^3 \simeq 0.104$ from the phase space.

\subsubsection{Constraints on the Higgs signal strengths}\label{ssSec:HiggsSignalStrengths}
Constraints on the Higgs signal strengths are imposed by using the $\kappa$ formalism \cite{LHCHiggsCrossSectionWorkingGroup:2013rie}. Given a production mechanism $i$ with cross section $\sigma_i$ or a decay process $i$ with width $\Gamma_i$, the parameter $\kappa_i$ is defined such that 
\begin{equation}\label{eq:Defkappa}
  \kappa_i^2 = \frac{\sigma_i}{\sigma_i^{\text{SM}}} \quad \text{or} \quad \kappa_i^2 = \frac{\Gamma_i}{\Gamma_i^{\text{SM}}},
\end{equation}
where $\sigma_i^{\text{SM}}$ and $\Gamma_i^{\text{SM}}$ are the corresponding SM quantities. The only two Higgs couplings affected at leading order are those to $AA$ and $AZ$. The corresponding $\kappa$'s are
\begin{equation}\label{eq:CaseIKappaI}
  \kappa_{AA}^2 = \frac{|S^{h\to AA}|^2 + |\tilde{S}^{h\to AA}|^2}{|S^{h\to AA}_\text{SM}|^2 + |\tilde{S}^{h\to AA}_\text{SM}|^2},\quad
  \kappa_{ZA}^2 = \frac{|S^{h\to ZA}|^2 + |\tilde{S}^{h\to ZA}|^2}{|S^{h\to ZA}_\text{SM}|^2 + |\tilde{S}^{h\to ZA}_\text{SM}|^2},
\end{equation}
where $\tilde{S}^{h\to AA}_\text{SM}$ and $\tilde{S}^{h\to ZA}_\text{SM}$ are both zero at leading order. The invisible ($A'A'$) and semi-invisible ($A'A$ and $A'Z$) decays of the Higgs boson are accounted for by rescaling the signal strengths. The experimental input is the Higgs signal strength measurements of Ref.~\cite{CMS:2022dwd} by CMS and Ref.~\cite{ATLAS:2022vkf} by ATLAS. These studies provide the measurements, uncertainties and correlations necessary to produce our own $\chi^2$ fit. The two searches are assumed to be uncorrelated.

\subsection{Oblique parameters}\label{sSec:ObliqueParameters}
We present in this section all contributions to the oblique parameters \cite{Peskin:1990zt}. Define
\begin{equation}\label{eq:ObliqueGeneral}
  \begin{aligned}
    &\alpha T_\phi(m, C, D, F, G, H^1, H^2)         = \\
    &  \frac{1}{32\pi^2 m_W^2}\Biggl[\sum_i \left(G_{ii} - c_W^2 D_{ii}\right)F_1(m_i) + \sum_{i, j}\left(|F_{ij}|^2 + |H^1_{ij}|^2 + |H^2_{ij}|^2- c_W^2 |C_{ij}|^2\right) F_2(m_i, m_j)\Biggr],\\
    &\alpha T_\psi(m, B^L, B^R, F^L, F^R) = \frac{1}{32\pi^2 m_W^2}\Biggl[(|F^L_{ij}|^2 + |F^R_{ij}|^2)F_3(m_i, m_j) + 2\text{Re}(F^L_{ij}F^{R\ast}_{ij})F_4(m_i, m_j)\\
    &                                       \hspace{4.5cm} - c_W^2\left[(|B^L_{ij}|^2 + |B^R_{ij}|^2)F_3(m_i, m_j) + 2\text{Re}(B^L_{ij}B^{R\ast}_{ij})F_4(m_i, m_j)\right] \Biggr],\\
    &\alpha S_\phi(m, A, C)               = \frac{s_W^2 c_W^2}{8 \pi^2}\sum_{i, j}\left(|C_{ij}|^2 - \frac{c_W^2 - s_W^2}{c_W s_W} C_{ij}A_{ji} - A_{ij}^2\right)F_5(m_i, m_j),\\
    &\alpha S_\psi(m, A, B^L, B^R)        = \frac{s_W^2 c_W^2}{8 \pi^2}\sum_{i, j}\Biggl[(|B^L_{ij}|^2 + |B^R_{ij}|^2)F_6(m_i, m_j) + 2\text{Re}(B^L_{ij}B^{R\ast}_{ij})F_7(m_i, m_j)\\
    &                                       \hspace{3.3cm}- \left(\frac{c_W^2 - s_W^2}{c_W s_W}(B^L_{ij} + B^R_{ij})A_{ij} + 2|A_{ij}|^2\right)\left[F_6(m_i, m_j) + F_7(m_i, m_j)\right] \Biggr],
  \end{aligned}
\end{equation}
where
\begin{equation}\label{eq:F1F2F3F4F5}
  \begin{aligned}
    F_1(m_i)      &= -4 m_i^2 (1 - \ln m_i^2),\\
    F_2(m_i, m_j) &= (m_i^2 + m_j^2)\left(3 - \frac{2(m_i^4 \ln m_i^2 - m_j^4 \ln m_j^2)}{m_i^4 - m_j^4}\right),\\
    F_3(m_i, m_j) &= (m_i^2 + m_j^2)\left(1 - \frac{2(m_i^4\ln m_i^2 - m_j^4 \ln m_j^2)}{m_i^4 - m_j^4}\right),\\
    F_4(m_i, m_j) &= -4 m_i m_j \left(1 - \frac{(m_i^2 \ln m_i^2  - m_j ^2\ln m_j^2)}{m_i^2 - m_j^2}\right),\\
    F_5(m_i, m_j) &= -\frac{5m_i^4 - 22m_i^2 m_j^2 + 5m_j^4}{9(m_i^2 - m_j^2)^2} + \frac{2\left[m_i^4(m_i^2 -3 m_j^2)\ln m_i^2 - m_j^4(m_j^2 - 3 m_i^2)\ln m_j^2\right]}{3(m_i^2 - m_j^2)^3},\\
    F_6(m_i, m_j) &= -\frac{4(m_i^4 - 8m_i^2 m_j^2 + m_j^4)}{9(m_i^2 - m_j^2)^2} + \frac{4\left[m_i^4(m_i^2 - 3m_j^2)\ln m_i^2 - m_j^4(m_j^2 - 3m_i^2)\ln m_j^2\right]}{3(m_i^2 - m_j^2)^3},\\
    F_7(m_i, m_j) &= -2m_i m_j\frac{(m_i^2 + m_j^2)}{(m_i^2 - m_j^2)^2}\left(1  - \frac{2m_i^2 m_j^2}{m_i^4 - m_j^4}\ln \frac{m_i^2}{m_j^2}\right).
  \end{aligned}
\end{equation}
The oblique parameters are then conveniently given by
\begin{equation}\label{eq:Oblique}
  \begin{aligned}
    T =& \; T_\phi(\hat{m}_C, \hat{C}_C, \hat{D}_C, \hat{F}_C, \hat{G}_C, \hat{H}^1_C, \hat{H}^2_C) + \frac{1}{2}T_\phi(\hat{m}_R, \hat{C}_R, \hat{D}_R, 0, \hat{G}_R, 0, 0)\\
    &+ T_\phi(\hat{m}_{RC}, 0, 0, \hat{F}_{RC}, 0, 0, 0)
    +\;T_\psi(\hat{m}_D, \hat{B}_D^L, \hat{B}_D^R, \hat{F}_D^L, \hat{F}_D^R) + \frac{1}{2}T_\psi(\hat{m}_M, \hat{B}_M^L, \hat{B}_M^R, 0, 0)\\
    &+ T_\psi(\hat{m}_{MD}, 0, 0, \hat{F}_{MD}^L, \hat{F}_{MD}^R),\\
    S =  &\;S_\phi(\hat{m}_C, \hat{A}_C, \hat{C}_C) + \frac{1}{2}S_\phi(\hat{m}_R, 0, \hat{C}_R) + S_\psi(\hat{m}_D, \hat{A}_D, \hat{B}_D^L, \hat{B}_D^R) + \frac{1}{2}S_\psi(\hat{m}_M, 0, \hat{B}_M^L, \hat{B}_M^R),
  \end{aligned}
\end{equation}
where 
\begin{equation}\label{eq:DefTS}
  \begin{aligned}
    &  \hat{m}_{RC} = \begin{pmatrix} \hat{m}_R &  \hat{m}_C \end{pmatrix},\quad \hat{m}_{MD} = \begin{pmatrix} \hat{m}_M & \hat{m}_D \end{pmatrix},\\
    & \hat{F}_{RC} = \begin{pmatrix} 0 & \hat{F}^1_{RC} \\ \hat{F}^2_{RC} & 0 \end{pmatrix},\quad \hat{F}^{L/R}_{MD} = \begin{pmatrix} 0 & \hat{F}^{1L/R}_{MD} \\ \hat{F}^{2L/R}_{MD} & 0 \end{pmatrix}.
  \end{aligned}
\end{equation}
A $\chi^2$ fit is performed using \cite{Workman:2022ynf}
\begin{equation}\label{eq:STpdf}
  S = -0.01 \pm 0.07, \qquad T = 0.04 \pm 0.06,
\end{equation}
with a correlation of 0.92.

\subsection{Unitarity}\label{sSec:Unitarity}
Consider the amplitude $\mathcal M$ of a $2 \to 2$ scattering process. It can be decomposed in terms of the Legendre polynomials $P_\ell(\cos\theta)$ as
\begin{equation}\label{eq:PartialWaves}
  \mathcal{M} = 16\pi\sum_\ell (2\ell + 1) a_\ell P_\ell(\cos\theta).
\end{equation}
For complex scalars, real scalars, Dirac spinors and Majorana spinors, we will define matrices of $a_\ell$ coefficients called $a_\ell^{\text{mat}}$. Unitarity will then impose
\begin{equation}\label{eq:UnitarityLimita0}
  a_\ell^{\text{max}} \equiv 
  \text{max}\left(\left|\text{Re}\left(a_\ell^{\text{eig}}\right)\right|\right) \leq \frac{1}{2},
\end{equation}
where $a_\ell^{\text{eig}}$ is the list of eigenvalues of $a_\ell^{\text{mat}}$. In the presence of identical particles in the incoming or outgoing state, $a_0$ is multiplied by a factor of $1/\sqrt{2}$. The limit of very high energy is assumed, in which case we can work with gauge eigenstates, though the basis invariance of the final results will be apparent. We will consider scattering from two components of an inert multiplet to two Higgs bosons and take inspiration from Ref.~\cite{Hally:2012pu}.

As a side note, we mention that the treatment of the fermion case differs from that of Refs.~\cite{Beauchesne:2022fet, Beauchesne:2023bcy}. Although the treatment of these references is very convenient for simpler cases, its generalization to more general cases can prove cumbersome.

\noindent{\it Complex scalars}: Consider the basis of field pairs
\begin{equation}\label{eq:BasisCS}
  hh, \phi^{C\dagger}_1\phi^C_1, \phi^{C\dagger}_1\phi^C_2, ..., \phi^{C\dagger}_1\phi^C_n, \phi^{C\dagger}_2\phi^C_1, \phi^{C\dagger}_2\phi^C_2, ..., \phi^{C\dagger}_2\phi^C_n,...
  \phi^{C\dagger}_n\phi^C_1, \phi^{C\dagger}_n\phi^C_2, ..., \phi^{C\dagger}_n\phi^C_n,
\end{equation}
where $n$ is the number of complex scalars. The $a_0^{\text{mat}}$ matrix for scattering from one pair to another is
\begin{equation}\label{eq:a0Complex}
  a_0^{\text{mat}} = -\frac{1}{16\sqrt{2} \pi}\begin{pmatrix} 0 & v^C \\ v^{C\dagger} & 0 \end{pmatrix},
\end{equation}
where
\begin{equation}\label{eq:vC}
  v^C = \begin{pmatrix}\lambda^C_{11} & \lambda^C_{12} & ... & \lambda^C_{1n} & \lambda^C_{21} & \lambda^C_{22} & ... \lambda^C_{2n} & ... & \lambda^C_{n1} & \lambda^C_{n2} & ... & \lambda^C_{nn} \end{pmatrix}.
\end{equation}
In practice, this simply gives
\begin{equation}\label{eq:UnitarityCS}
  a_0^{\text{max}} = \frac{\sqrt{\text{Tr}[(\lambda^C)^2]}}{16\sqrt{2}\pi}.
\end{equation}

\noindent{\it Real scalars}: Consider the basis of field pairs
\begin{equation}\label{eq:BasisRS}
  hh, \phi^R_1\phi^R_1, \phi^R_1\phi^R_2, ..., \phi^R_1\phi^R_n, \phi^R_2\phi^R_2, \phi^R_2\phi^R_3, ..., \phi^R_2\phi^R_n,...,\phi^R_n\phi^R_n,
\end{equation}
where $n$ is the number of real scalars. The $a_0^{\text{mat}}$ matrix for scattering from one pair to another is
\begin{equation}\label{eq:a0Real}
  a_0^{\text{mat}} = -\frac{1}{16\sqrt{2}\pi}\begin{pmatrix} 0 & v^R \\ v^{R \dagger} & 0 \end{pmatrix},
\end{equation}
where
\begin{equation}\label{eq:vR}
  v^R = \begin{pmatrix} \frac{\lambda^R_{11}}{\sqrt{2}} & \lambda^R_{12} & ... & \lambda^R_{1n} & \frac{\lambda^R_{22}}{\sqrt{2}} & \lambda^R_{23} & ... \lambda^R_{2n} & ... & \frac{\lambda^R_{nn}}{\sqrt{2}} \end{pmatrix},
\end{equation}
In practice, this simply gives
\begin{equation}\label{eq:UnitarityRS}
  a_0^{\text{max}} = \frac{\sqrt{\text{Tr}[(\lambda^R)^2]}}{32\pi}.
\end{equation}

\noindent{\it Dirac fermions}: Consider the basis of field pairs
\begin{equation}\label{eq:BasisDirac}
  hh, \bar{\psi}^D_1\psi^D_1, \bar{\psi}^D_1\psi^D_2, ..., \bar{\psi}^D_1\psi^D_n, \bar{\psi}^D_2\psi^D_1, \bar{\psi}^D_2\psi^D_2, ..., \bar{\psi}^D_2\psi^D_n,... \bar{\psi}^D_n\psi^D_1, \bar{\psi}^D_n\psi^D_2, ..., \bar{\psi}^D_n\psi^D_n,
\end{equation}
where $n$ is the number of Dirac fermions. The $a_0$ coefficients for $\bar{\psi}_i^D \psi_j^D \to h h$ are null. The $a_1^{\text{mat}}$ matrix for scattering from one pair to another is
\begin{equation}\label{eq:a0Dirac}
  a_1^{\text{mat}} = -\frac{1}{16\sqrt{2}}\begin{pmatrix} 0 & v^D \\ v^{D\dagger} & 0 \end{pmatrix},
\end{equation}
where
\begin{equation}\label{eq:vD}
  v^D = \begin{pmatrix} A^D_{11}  & A^D_{12} & ... & A^D_{1n} &  A^D_{21}  & A^D_{22} & ... & A^D_{2n} & A^D_{n1}  & A^D_{n2} & ... & A^D_{nn} \end{pmatrix},
\end{equation}
with
\begin{equation}\label{eq:AD}
  A^D_{ij} = \begin{pmatrix} \sum_k (\Omega^R_D)_{ik}(\Omega^L_D)_{kj} & \sum_k (\Omega^L_D)_{ik}(\Omega^R_D)_{kj} \end{pmatrix},
\end{equation}
in the helicity basis $(\uparrow\downarrow, \downarrow\uparrow)$. In practice, this simply gives
\begin{equation}\label{eq:UnitarityDF}
  a_1^{\text{max}} = \frac{\sqrt{\text{Tr}[(\Omega_D^L \Omega_D^R)^2]}}{16}.
\end{equation}

\noindent{\it Majorana fermions}: Consider the basis of field pairs
\begin{equation}\label{eq:BasisMajorana}
  hh, \psi^M_1\psi^M_1, \psi^M_1\psi^M_2, ..., \psi^M_1\psi^M_n, \psi^M_2\psi^M_2, \psi^M_2\psi^M_3, ..., \psi^M_2\psi^M_n,..., \psi^M_n\psi^M_n,
\end{equation}
where $n$ is the number of Majorana fermions. The $a_0$ coefficients for $\psi_i^M \psi_j^M \to h h$ are null. The $a_1^{\text{mat}}$ matrix for scattering from one pair to another is
\begin{equation}\label{eq:a1Majorana}
  a_1^{\text{mat}} = -\frac{1}{16\sqrt{2}}\begin{pmatrix} 0 & v^M \\ v^{M\dagger} & 0 \end{pmatrix},
\end{equation}
where
\begin{equation}\label{eq:vM}
   v^M = \begin{pmatrix} \frac{A^M_{11}}{\sqrt{2}}  & A^M_{12} & ... & A^M_{1n} &  \frac{A^M_{22}}{\sqrt{2}}  & A^M_{23} & ... & A^M_{2n} & ... & \frac{A^M_{nn}}{\sqrt{2}} \end{pmatrix},
\end{equation}
with
\begin{equation}\label{eq:AM}
  A^M_{ij} = \begin{pmatrix} \sum_k (\Omega^R_M)_{ik}(\Omega^L_M)_{kj} & \sum_k (\Omega^L_M)_{ik}(\Omega^R_M)_{kj} \end{pmatrix},
\end{equation}
in the helicity basis $(\uparrow\downarrow, \downarrow\uparrow)$. In practice, this simply gives
\begin{equation}\label{eq:UnitarityMF}
    a_1^{\text{max}} = \frac{\sqrt{\text{Tr}[(\Omega_M^L \Omega_M^R)^2]}}{16\sqrt{2}},
\end{equation}
where we have assumed that $\Omega^R_M$ and $\Omega^L_M$ are constructed such that they are symmetric, which can always be performed.

\subsection{Triple Higgs coupling}\label{sSec:TripleHiggsCoupling}
With our formalism, the triple Higgs coupling can easily be computed using the Coleman-Weinberg potential \cite{Coleman:1973jx}. In the $\overline{MS}$ renormalization scheme, it is given by
\begin{equation}\label{eq:CWpotential}
  \begin{aligned}
    V_{CW} = \frac{1}{64\pi^2} \biggl[     &\sum_i(\hat{m}^4_R)_{ii} \left(\ln\left(\frac{(\hat{m}^2_R)_{ii}}{Q^2}\right) -\frac{3}{2}\right) 
                                        + 2 \sum_i(\hat{m}^4_C)_{ii} \left(\ln\left(\frac{(\hat{m}^2_C)_{ii}}{Q^2}\right) -\frac{3}{2}\right) \\
                                        - 2&\sum_i(\hat{m}^4_M)_{ii} \left(\ln\left(\frac{(\hat{m}^2_M)_{ii}}{Q^2}\right) -\frac{3}{2}\right) 
                                        - 4 \sum_i(\hat{m}^4_D)_{ii} \left(\ln\left(\frac{(\hat{m}^2_D)_{ii}}{Q^2}\right) -\frac{3}{2}\right) \biggr], 
  \end{aligned}
\end{equation}
where the different masses are understood to be the field-dependent masses and $Q$ is some scale that cancels in the final result. Requesting to reproduce the correct mass and VEV of the Higgs, the triple Higgs coupling $\lambda_{hhh}$ is given by
\begin{equation}\label{eq:HiggsTripleCoupling}
  \frac{\lambda_{hhh}}{3!} = \frac{m_h^2}{2v} + \left[\left( \frac{1}{2v^2}\frac{\partial}{\partial h} - \frac{1}{2v}\frac{\partial^2}{\partial h^2} + \frac{1}{3!}\frac{\partial^3}{\partial h^3}\right)V_{CW}\right]\Biggr|_{h=0},
\end{equation}
where we use the convention that $v\approx 246$~GeV.

\section{Results}\label{Sec:Results}
We present in this section the potential contributions of inert multiplets to $\text{BR}(h \to Z A^{(\prime)})$ and the triple Higgs coupling. We first discuss our scanning procedure and then present results in the presence of either one or multiple interaction terms.

\subsection{Scanning procedure}\label{sSec:Scanning}
For each benchmark model considered, the entire parameter space is scanned using a Markov chain with the Metropolis-Hasting algorithm. To increase the density of points toward the limits and therefore obtain faster convergence, a non-flat prior is chosen proportional to
\begin{equation}\label{eq:Prior}
  \left(\frac{\delta\lambda_{hhh}}{a}\right)^2 + \left(\frac{\delta\text{BR}(h \to A Z)}{b}\right)^2 + \left(\frac{\text{BR}(h \to A' Z)}{c}\right)^2
\end{equation}
where we set $a=b=0.1$ and $c=0.001$. The relative deviation of the triple Higgs coupling from its SM value is defined as
\begin{equation}\label{eq:deltahhh}
  \delta\lambda_{hhh} = \frac{\lambda_{hhh} -\lambda_{hhh}^{\text{SM}}}{\lambda_{hhh}^{\text{SM}}}.
\end{equation}
The relative deviation of $\text{BR}(h \to A Z)$ from its SM value is defined as
\begin{equation}\label{eq:deltaBAZ}
  \delta\text{BR}(h \to A Z) = \frac{\text{BR}(h \to A Z) - \text{BR}(h \to A Z)_{\text{SM}}}{\text{BR}(h \to A Z)_{\text{SM}}}.
\end{equation}
We have verified that the Markov chain converges to the same results irrespective of the choice of prior. To keep the number of figures reasonable, the results are finely binned and presented as histograms. When multiple contractions of $SU(2)_L$ indices are possible, all contractions are considered. For scalar multiplets, points are dismissed in the presence of tachyons, as the scalars acquiring VEVs fall outside the scope of this work and can even potentially break electromagnetism.\footnote{We mention that Higgs vacuum instability could also be a problem, for example because the fermionic multiplets tend to make the Higgs quartic run more negative. However, bounds on vacuum stability would imply the assumption that no new physics exists between the electroweak scale and the instability scale. Since we cannot guarantee this assumption is respected, we do not impose such bounds. In the same vein, the existence of additional minima that break electromagnetism could potentially be problematic. However, such bounds would depend on the full potential. The additional terms could address the stability issues while leaving our observables unchanged at leading order. Because of this, we do not consider this issue any further.} Points that contain charged particles below 100~GeV are also rejected, as such particles are ruled out by LEP \cite{LEP1, LEP2}. As discussed in Refs.~\cite{Beauchesne:2022fet, Beauchesne:2023bcy}, constraints from the Higgs signal strengths can be partially avoided by having purely imaginary $\hat{\Omega}^{L/R}_D$ couplings. This however leads to a strongly excluded contribution to the electron electric dipole moment, which forces $\hat{\Omega}^{L/R}_D$ to be almost purely real. In practice, the results are equivalent to taking $A_{L/R}$ to be real. For the sake of simplicity and to speed up computations, we will therefore take $A_{L/R}$ real. For complex scalars, unitarity imposes a constraint on the couplings to dark photons of \cite{Hally:2012pu}
\begin{equation}\label{eq:Unitarity}
  \sum_i (A'_{C})_{ii}^4 < 16 \pi^2.
\end{equation}
For Dirac fermions, we take the less restrictive $\text{max}(|A'_D|_{ii}) < \sqrt{4\pi}$. As a validation procedure, we have verified that all divergences in the loop decay widths and oblique parameters cancel for all points of every benchmark. Finally, as pointed out in Refs.~\cite{Beauchesne:2022fet, Beauchesne:2023bcy}, the Higgs signal strengths are considerably insensitive to the scenario that the BSM contributions to $S^{h \to A A}$ ($S^{h \to A Z}$) are about $-2S^{h \to A A}_{\text{SM}}$ ($-2S^{h \to A Z}_{\text{SM}}$), since this simply changes the sign of $S^{h \to A A}$ ($S^{h \to A Z}$) which the Higgs signal strengths are insensitive to. This would however require tremendous tuning, if even possible, and we will ignore such points.

\subsection{One interaction term}\label{sSec:OneVertexResults}
We first present results in the presence of a single interaction term of the types presented in Sec.~\ref{Sec:Vertices}. The chosen benchmarks are presented in Tables~\ref{table:BenchmarksI} and \ref{table:BenchmarksII}. Each entry contains the relevant fields and their interaction Lagrangian as defined in Sec.~\ref{Sec:Vertices}. All bounds are at $95\%$ confidence level.

\begin{table}[t!]
\centering
\begin{tabular}{m{1cm} m{1cm} m{2.6cm} m{1.7cm} m{2cm}} 
 \hline
 Name & Fields                          & Gauge numbers                                                          & Type              & Lagrangian            \\
 \hline
 FA   & $\psi_1$ \hspace{2cm} $\psi_2$  & $(\mathbf{2}, 1/2, 1)$ $(\mathbf{1},    0, 1)$ & Complex     Complex    & $F(\psi_1, \psi_2)$   \\ 
 \hline
 FB   & $\psi_1$ \hspace{2cm} $\psi_2$  & $(\mathbf{1},   0, 1)$ $(\mathbf{2}, -1/2, 1)$ & Complex     Complex    & $F(\psi_1, \psi_2)$   \\ 
 \hline
 FC   & $\psi_1$ \hspace{2cm} $\psi_2$  & $(\mathbf{3},   0, 1)$ $(\mathbf{2}, -1/2, 1)$ & Complex     Complex    & $F(\psi_1, \psi_2)$   \\ 
 \hline
 FD   & $\psi_1$ \hspace{2cm} $\psi_2$  & $(\mathbf{2}, 1/2, 1)$ $(\mathbf{3},    0, 1)$ & Complex     Complex    & $F(\psi_1, \psi_2)$   \\ 
 \hline
 FE   & $\psi_1$ \hspace{2cm} $\psi_2$  & $(\mathbf{2}, 1/2, 0)$ $(\mathbf{1},    0, 0)$ & Complex\;     Real & $F(\psi_1, \psi_2)$   \\ 
 \hline
 FF   & $\psi_1$ \hspace{2cm} $\psi_2$  & $(\mathbf{1},   0, 0)$ $(\mathbf{2},  -1/2, 0)$ & Real  Complex    & $F(\psi_1, \psi_2)$   \\ 
 \hline
  FG   & $\psi_1$ \hspace{2cm} $\psi_2$  & $(\mathbf{3},   0, 0)$ $(\mathbf{2},  -1/2, 0)$ & Real  Complex    & $F(\psi_1, \psi_2)$   \\ 
 \hline
  FH   & $\psi_1$ \hspace{2cm} $\psi_2$  & $(\mathbf{2}, 1/2, 0)$ $(\mathbf{3},   
  0, 0)$ & Complex  Real    & $F(\psi_1, \psi_2)$   \\ 
 \hline
 S1A  & $\phi_1$ \hspace{2cm} $\phi_2$  & $(\mathbf{2}, 1/2, 1)$ $(\mathbf{1},    0, 1)$ & Complex   Complex  & $S_1(\phi_1, \phi_2)$ \\ 
 \hline
 S1B  & $\phi_1$ \hspace{2cm} $\phi_2$  & $(\mathbf{1},   0, 1)$ $(\mathbf{2}, -1/2, 1)$ & Complex   Complex  & $S_1(\phi_1, \phi_2)$ \\ 
 \hline
 S1C  & $\phi_1$ \hspace{2cm} $\phi_2$  & $(\mathbf{3},   0, 1)$ $(\mathbf{2}, -1/2, 1)$ & Complex   Complex  & $S_1(\phi_1, \phi_2)$ \\ 
 \hline
 S1D  & $\phi_1$ \hspace{2cm} $\phi_2$  & $(\mathbf{2}, 1/2, 1)$ $(\mathbf{3},    0, 1)$ & Complex   Complex  & $S_1(\phi_1, \phi_2)$ \\ 
 \hline
 S1E  & $\phi_1$ \hspace{2cm} $\phi_2$  & $(\mathbf{2}, 1/2, 0)$ $(\mathbf{1},    0, 0)$ & Complex   Real     & $S_1(\phi_1, \phi_2)$ \\ 
 \hline
 S1F  & $\phi_1$ \hspace{2cm} $\phi_2$  & $(\mathbf{1},   0, 0)$ $(\mathbf{2}, -1/2, 0)$ & Real      Complex  & $S_1(\phi_1, \phi_2)$ \\ 
 \hline
 S1G  & $\phi_1$ \hspace{2cm} $\phi_2$  & $(\mathbf{3},   0, 0)$ $(\mathbf{2}, -1/2, 0)$ & Real      Complex  & $S_1(\phi_1, \phi_2)$ \\ 
 \hline
 S1H  & $\phi_1$ \hspace{2cm} $\phi_2$  & $(\mathbf{2}, 1/2, 0)$ $(\mathbf{3},    0, 0)$ & Complex   Real     & $S_1(\phi_1, \phi_2)$ \\ 
 \hline
 S2A  & $\phi$                          & $(\mathbf{1},   1, 1)$                                     & Complex            & $S_2(\phi)$           \\ 
 \hline
 S2B  & $\phi$                          & $(\mathbf{1},   2, 1)$                                     & Complex            & $S_2(\phi)$           \\ 
 \hline
 S2C  & $\phi$                          & $(\mathbf{2}, 1/2, 1)$                                     & Complex            & $S_2(\phi)$           \\ 
 \hline
 S2D  & $\phi$                          & $(\mathbf{2}, 3/2, 1)$                                     & Complex            & $S_2(\phi)$           \\ 
 \hline
\end{tabular}
\caption{Different benchmarks for the single interaction term case. The type represents whether the multiplet is real or complex. The gauge numbers are in the format $(SU(2)_L, U(1)_Y, U(1)')$.}
\label{table:BenchmarksI}
\end{table}

\begin{table}[t!]
\centering
\begin{tabular}{m{1cm} m{1cm} m{2.6cm} m{1.6cm} m{2cm}}
 \hline
 Name & Fields                          & Gauge numbers                                                          & Type              & Lagrangian            \\
 \hline
  S2E  & $\phi$                          & $(\mathbf{3},   0, 1)$                                     & Complex            & $S_2(\phi)$           \\ 
 \hline
  S2F  & $\phi$                          & $(\mathbf{3},   1, 1)$                                     & Complex               & $S_2(\phi)$           \\ 
 \hline
  S2G  & $\phi$                          & $(\mathbf{1},   0, 0)$                                     & Real               & $S_2(\phi)$           \\ 
 \hline
  S2H  & $\phi$                          & $(\mathbf{3},   0, 0)$                                     & Real               & $S_2(\phi)$           \\ 
 \hline
  S3A  & $\phi_1$ \hspace{2cm} $\phi_2$  & $(\mathbf{1},   1, 1)$\; $(\mathbf{1},    1, 1)$ & Complex   Complex  & $S_3(\phi_1, \phi_2)$ \\ 
 \hline
  S3B  & $\phi_1$ \hspace{2cm} $\phi_2$  & $(\mathbf{1},   2, 1)$\; $(\mathbf{1},    2, 1)$ & Complex   Complex  & $S_3(\phi_1, \phi_2)$ \\ 
 \hline
 S3C  & $\phi_1$ \hspace{2cm} $\phi_2$  & $(\mathbf{2}, 1/2, 1)$\; $(\mathbf{2},  1/2, 1)$ & Complex   Complex  & $S_3(\phi_1, \phi_2)$ \\ 
 \hline
 S3D  & $\phi_1$ \hspace{2cm} $\phi_2$  & $(\mathbf{2}, 3/2, 1)$\; $(\mathbf{2},  3/2, 1)$ & Complex   Complex  & $S_3(\phi_1, \phi_2)$ \\ 
 \hline
 S3E  & $\phi_1$ \hspace{2cm} $\phi_2$  & $(\mathbf{3},   0, 1)$\; $(\mathbf{3},    0, 1)$ & Complex   Complex  & $S_3(\phi_1, \phi_2)$ \\ 
 \hline
 S3F  & $\phi_1$ \hspace{2cm} $\phi_2$  & $(\mathbf{3},   1, 1)$\; $(\mathbf{3},    1, 1)$ & Complex   Complex  & $S_3(\phi_1, \phi_2)$ \\ 
 \hline
 S3G  & $\phi_1$ \hspace{2cm} $\phi_2$  & $(\mathbf{1},   0, 0)$\; $(\mathbf{1},    0, 0)$ & Real\quad Real     & $S_3(\phi_1, \phi_2)$ \\ 
 \hline
 S3H  & $\phi_1$ \hspace{2cm} $\phi_2$  & $(\mathbf{3},   0, 0)$\; $(\mathbf{3},    0, 0)$ & Complex   Real     & $S_3(\phi_1, \phi_2)$ \\ 
 \hline
 S3I  & $\phi_1$ \hspace{2cm} $\phi_2$  & $(\mathbf{3},   0, 0)$\; $(\mathbf{3},    0, 0)$ & Real\quad Real     & $S_3(\phi_1, \phi_2)$ \\ 
 \hline
 S4A  & $\phi_1$ \hspace{2cm} $\phi_2$  & $(\mathbf{2},   1, 1)$\; $(\mathbf{2},    0, 1)$ & Complex   Complex  & $S_4(\phi_1, \phi_2)$ \\ 
 \hline
 S4B  & $\phi_1$ \hspace{2cm} $\phi_2$  & $(\mathbf{3},   1, 1)$\; $(\mathbf{1},    0, 1)$ & Complex   Complex  & $S_4(\phi_1, \phi_2)$ \\ 
 \hline
 S4C  & $\phi_1$ \hspace{2cm} $\phi_2$  & $(\mathbf{3},   1, 1)$\; $(\mathbf{3},    0, 1)$ & Complex   Complex  & $S_4(\phi_1, \phi_2)$ \\ 
 \hline
 S4D  & $\phi_1$ \hspace{2cm} $\phi_2$  & $(\mathbf{3},   1, 0)$\; $(\mathbf{1},    0, 0)$ & Complex   Real     & $S_4(\phi_1, \phi_2)$ \\ 
 \hline
 S4E  & $\phi_1$ \hspace{2cm} $\phi_2$  & $(\mathbf{3},   1, 0)$\; $(\mathbf{3},    0, 0)$ & Complex   Real     & $S_4(\phi_1, \phi_2)$ \\ 
 \hline
 S5A  & $\phi$                          & $(\mathbf{2}, 1/2, 0)$                                     & Complex & $S_5(\phi)$           \\ 
 \hline
 S5B  & $\phi$                          & $(\mathbf{4}, 1/2, 0)$                                     & Complex  & $S_5(\phi)$           \\ 
 \hline
 S6A  & $\phi$                          & $(\mathbf{2}, 0, 0)$                                     & Complex & $S_6(\phi)$           \\ 
 \hline
 S6B  & $\phi$                          & $(\mathbf{3}, 0, 0)$                                     & Complex & $S_6(\phi)$           \\ 
 \hline
 S6C  & $\phi$                          & $(\mathbf{3}, 0, 0)$                                     & Real  & $S_6(\phi)$           \\ 
 \hline
\end{tabular}
\caption{Table~\ref{table:BenchmarksI} continued}
\label{table:BenchmarksII}
\end{table}

\begin{figure*}[t!]
\begin{minipage}[b]{.97\textwidth}
\begin{center}
 \captionsetup[subfigure]{justification=centerlast}
 \begin{subfigure}{0.49\textwidth}
    \centering
    \caption{Fermion case}
    \includegraphics[width=1\textwidth]{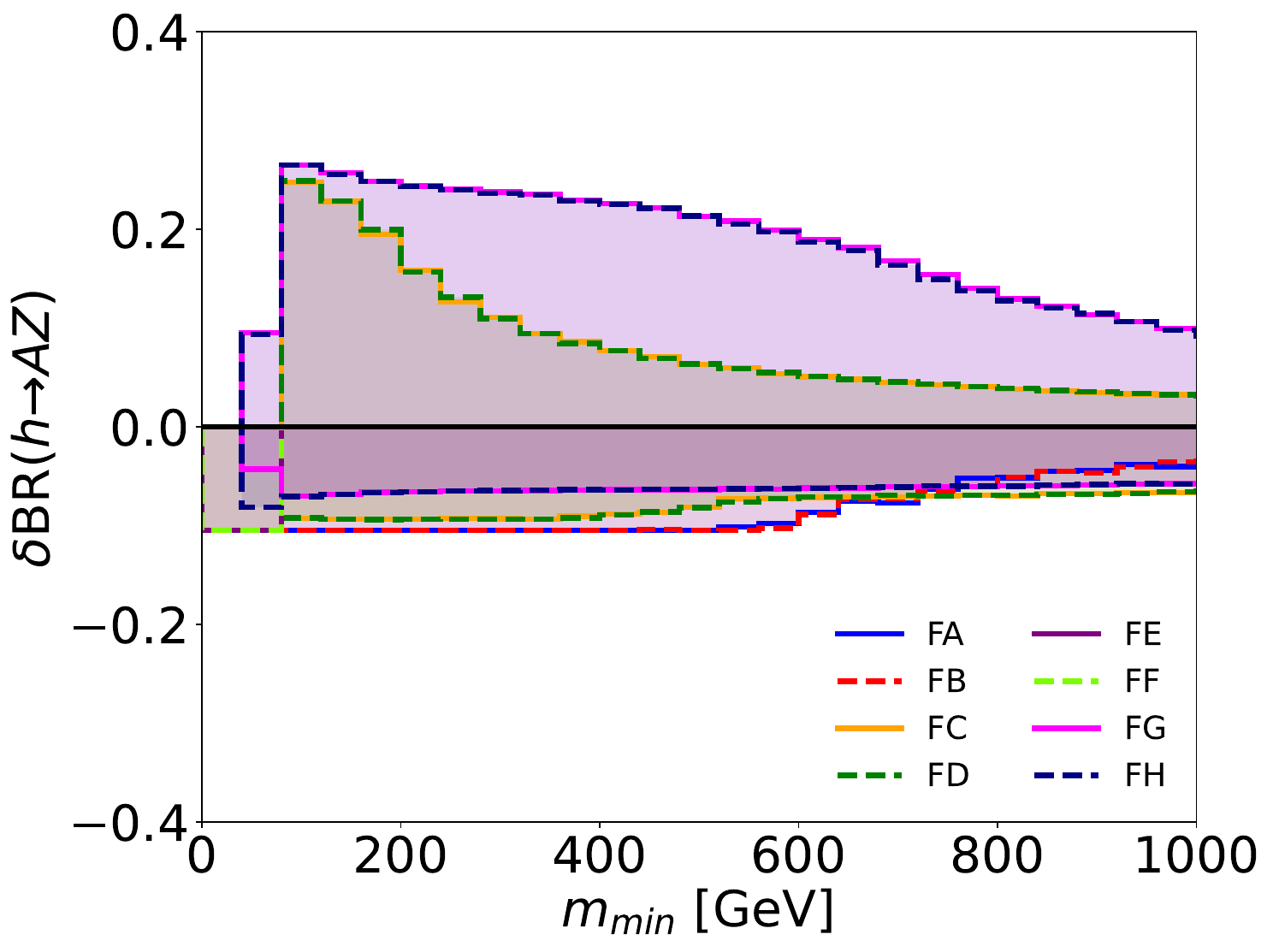}
    \label{fig:FBRAZ}
 \end{subfigure}
 \captionsetup[subfigure]{justification=centerlast}
 \begin{subfigure}{0.49\textwidth}
    \centering
    \caption{Scalar case I}
    \includegraphics[width=1\textwidth]{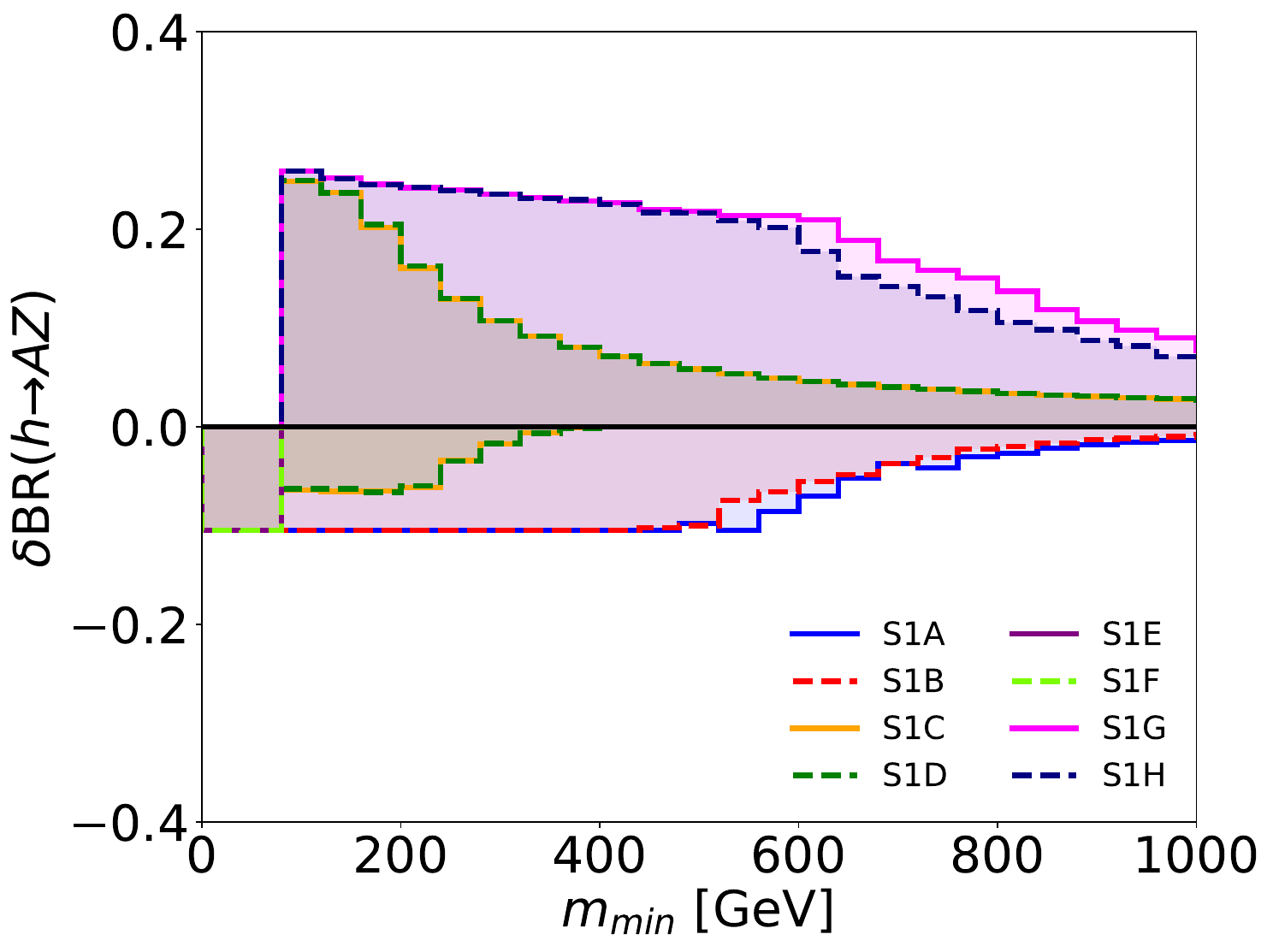}
    \label{fig:S1BRAZ}
 \end{subfigure}
 \captionsetup[subfigure]{justification=centerlast}
 \begin{subfigure}{0.49\textwidth}
    \centering
    \caption{Scalar case II}
    \includegraphics[width=1\textwidth]{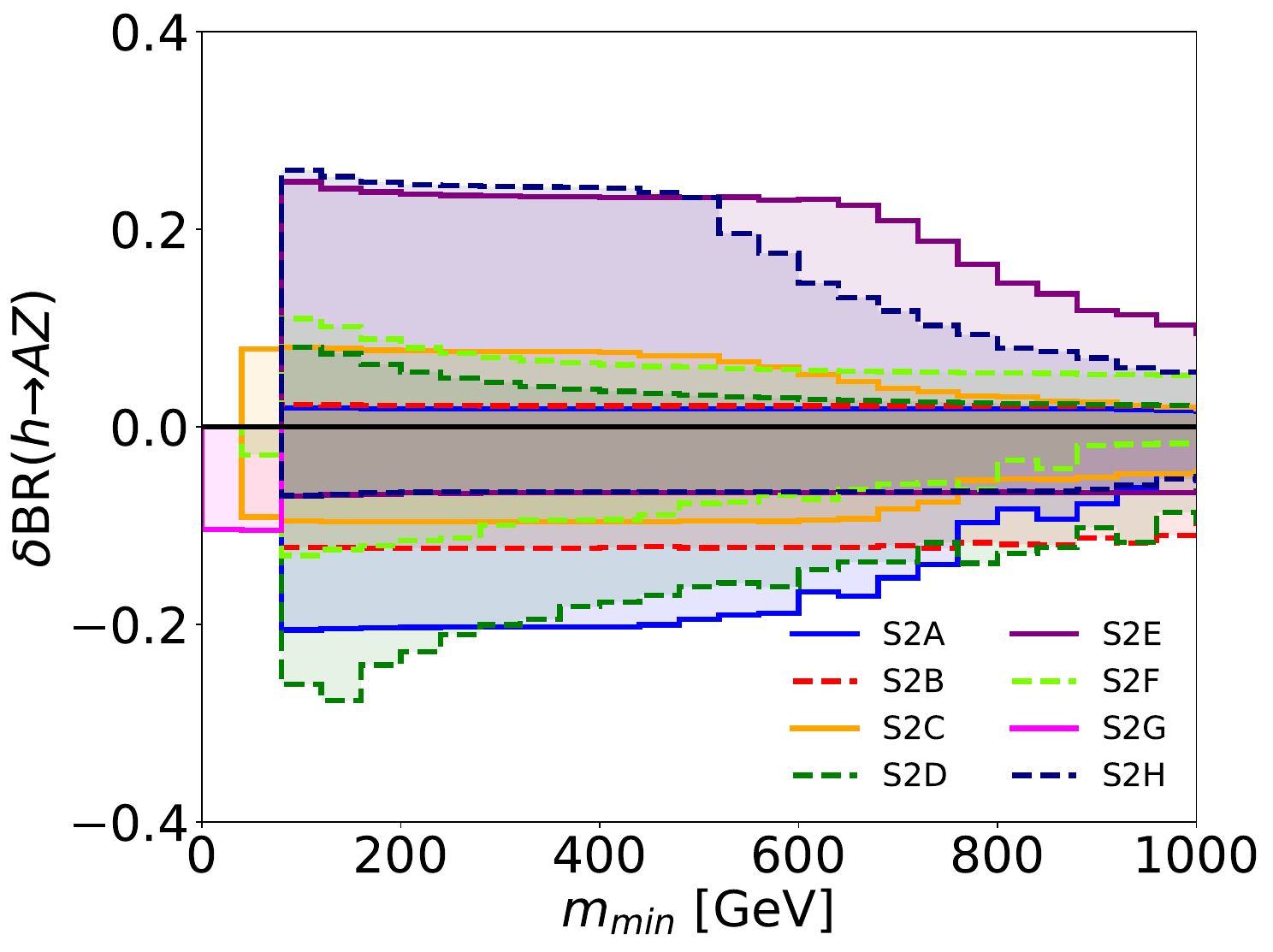}
    \label{fig:S2BRAZ}
 \end{subfigure}
 \captionsetup[subfigure]{justification=centerlast}
 \begin{subfigure}{0.49\textwidth}
    \centering
    \caption{Scalar case III}
    \includegraphics[width=1\textwidth]{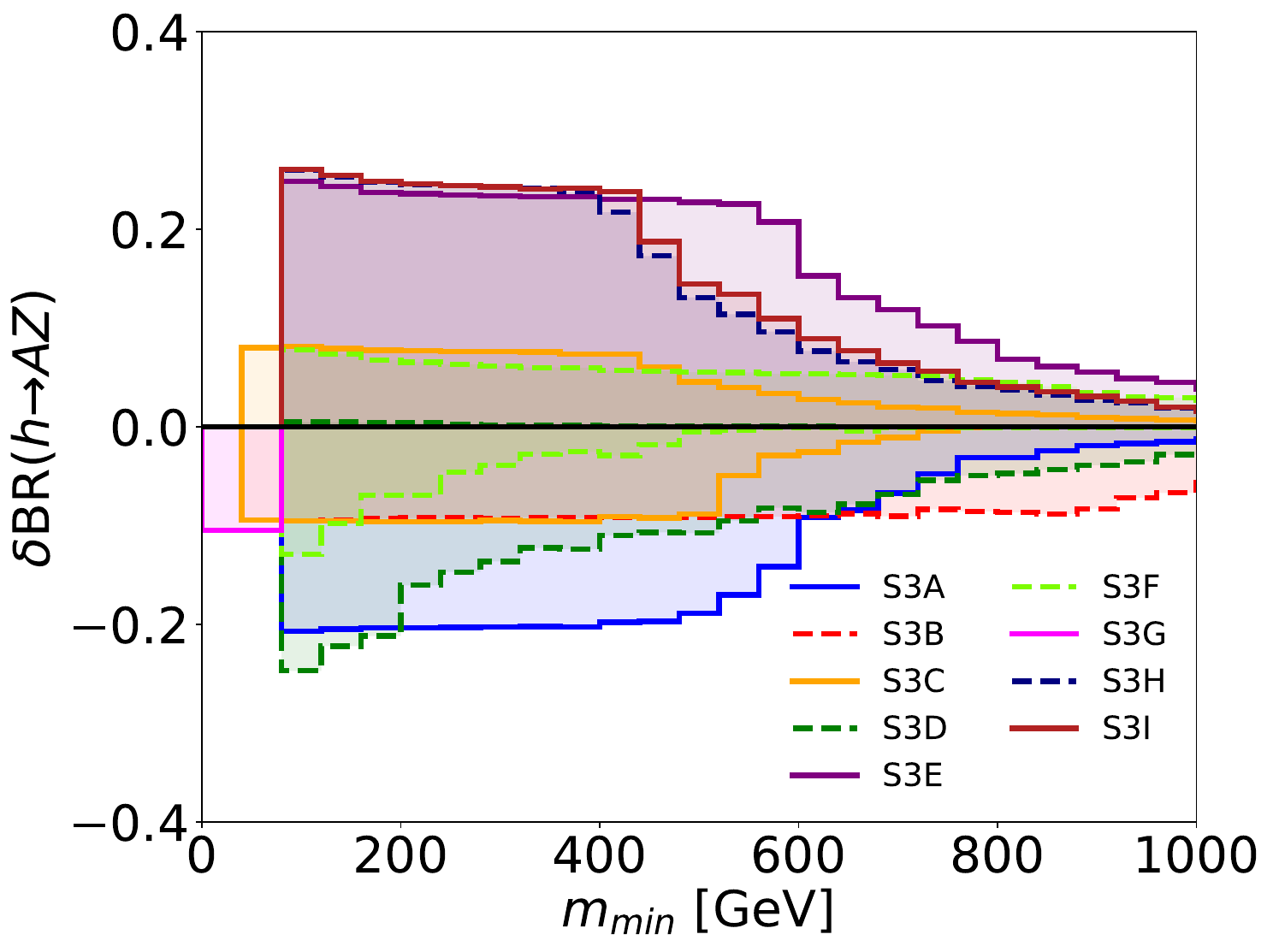}
    \label{fig:S3BRAZ}
 \end{subfigure}
  \captionsetup[subfigure]{justification=centerlast}
 \begin{subfigure}{0.49\textwidth}
    \centering
    \caption{Scalar case IV}
    \includegraphics[width=1\textwidth]{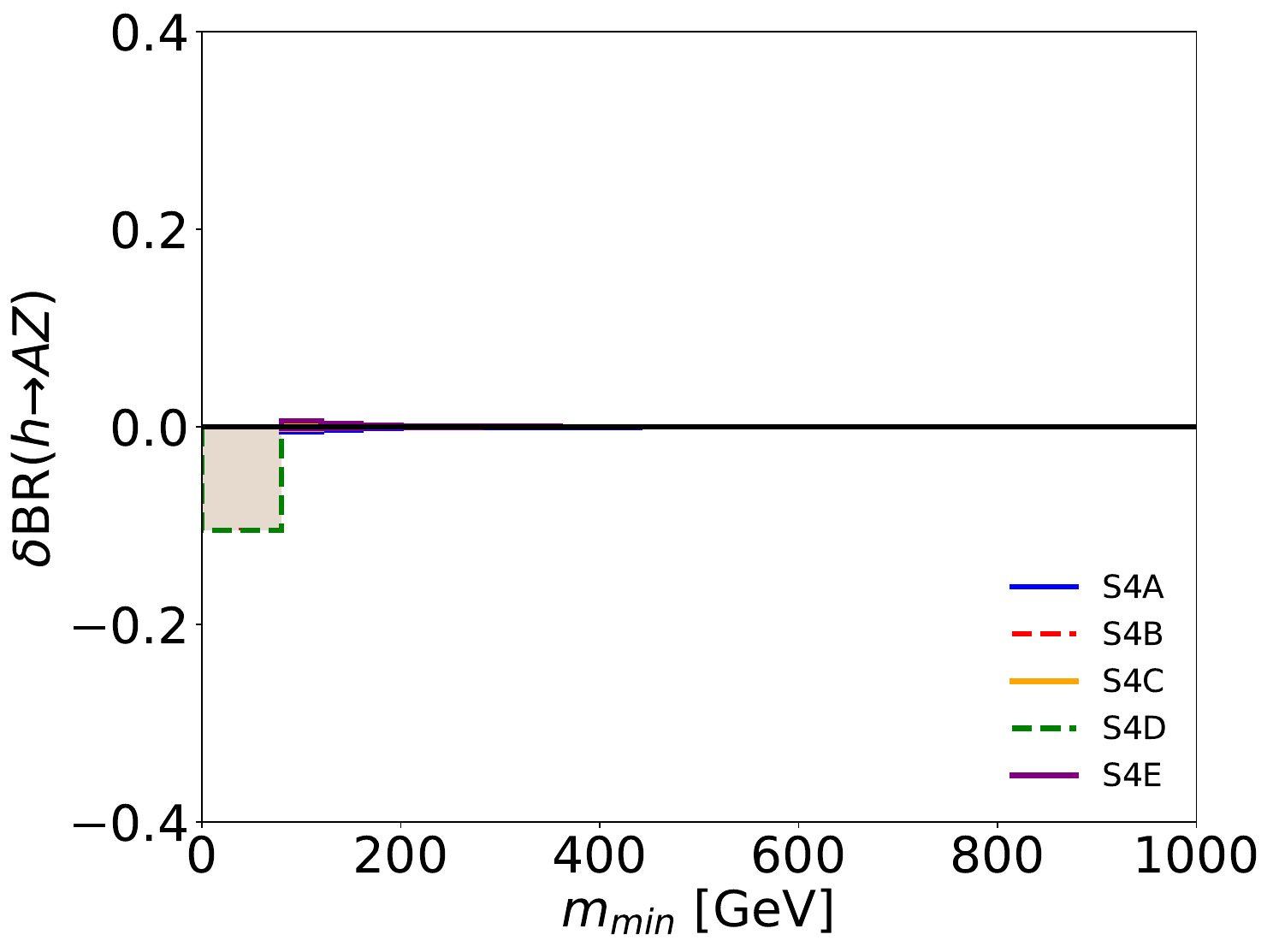}
    \label{fig:S4BRAZ}
 \end{subfigure}
  \captionsetup[subfigure]{justification=centerlast}
 \begin{subfigure}{0.49\textwidth}
    \centering
    \caption{Scalar case V + VI}
    \includegraphics[width=1\textwidth]{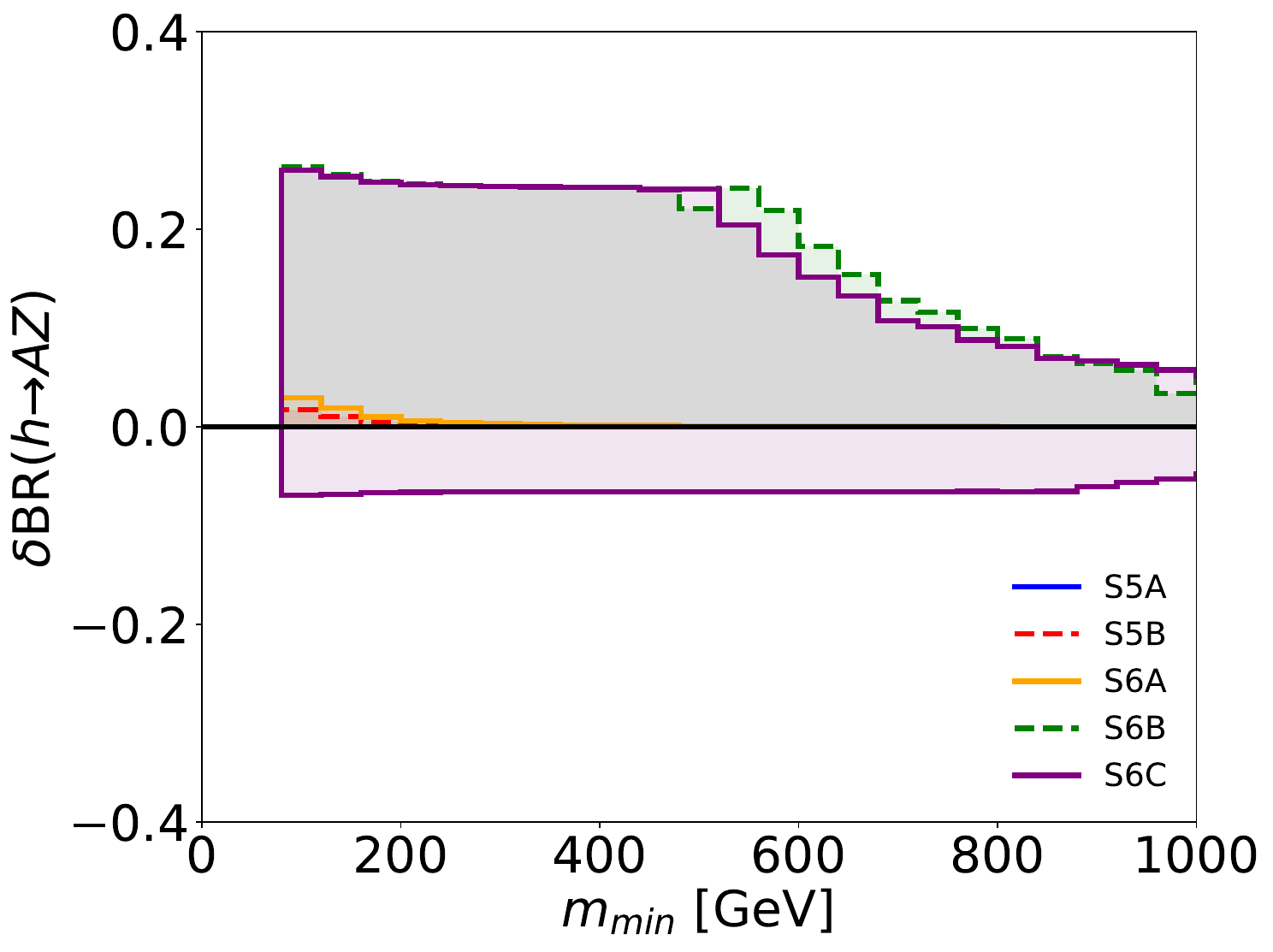}
    \label{fig:S5BRAZ}
 \end{subfigure}
\caption{Allowed range of $\delta \text{BR}(h \to AZ)$ for different cases of a single interaction term. See Tables~\ref{table:BenchmarksI} and~\ref{table:BenchmarksII} for a description of the benchmarks.}
\label{fig:BRAZ}
\end{center}
\end{minipage}
\end{figure*}

First, Fig.~\ref{fig:BRAZ} shows the relative deviation of $\text{BR}(h \to A Z)$ from its SM value as a function of the mass of the lightest new particles $m_{\text{min}}$. As can be seen, deviations of $\pm \mathcal{O}(20\%)$ are possible. The scalar cases IV and V are especially constrained and do not lead to any sizable contribution to this observable. This will be the case for all other observables. The only exception is if the decay of the Higgs boson to two new scalars is allowed, in which case $\text{BR}(h \to A Z)$ can be reduced. Many other benchmark models with a single interaction term have been analyzed, but none allowed for a significantly larger range of $\delta\text{BR}(h \to A Z)$.

\begin{figure*}[t!]
\begin{minipage}[b]{.97\textwidth}
\begin{center}
 \captionsetup[subfigure]{justification=centerlast}
 \begin{subfigure}{0.49\textwidth}
    \centering
    \caption{Fermion case}
    \includegraphics[width=1\textwidth]{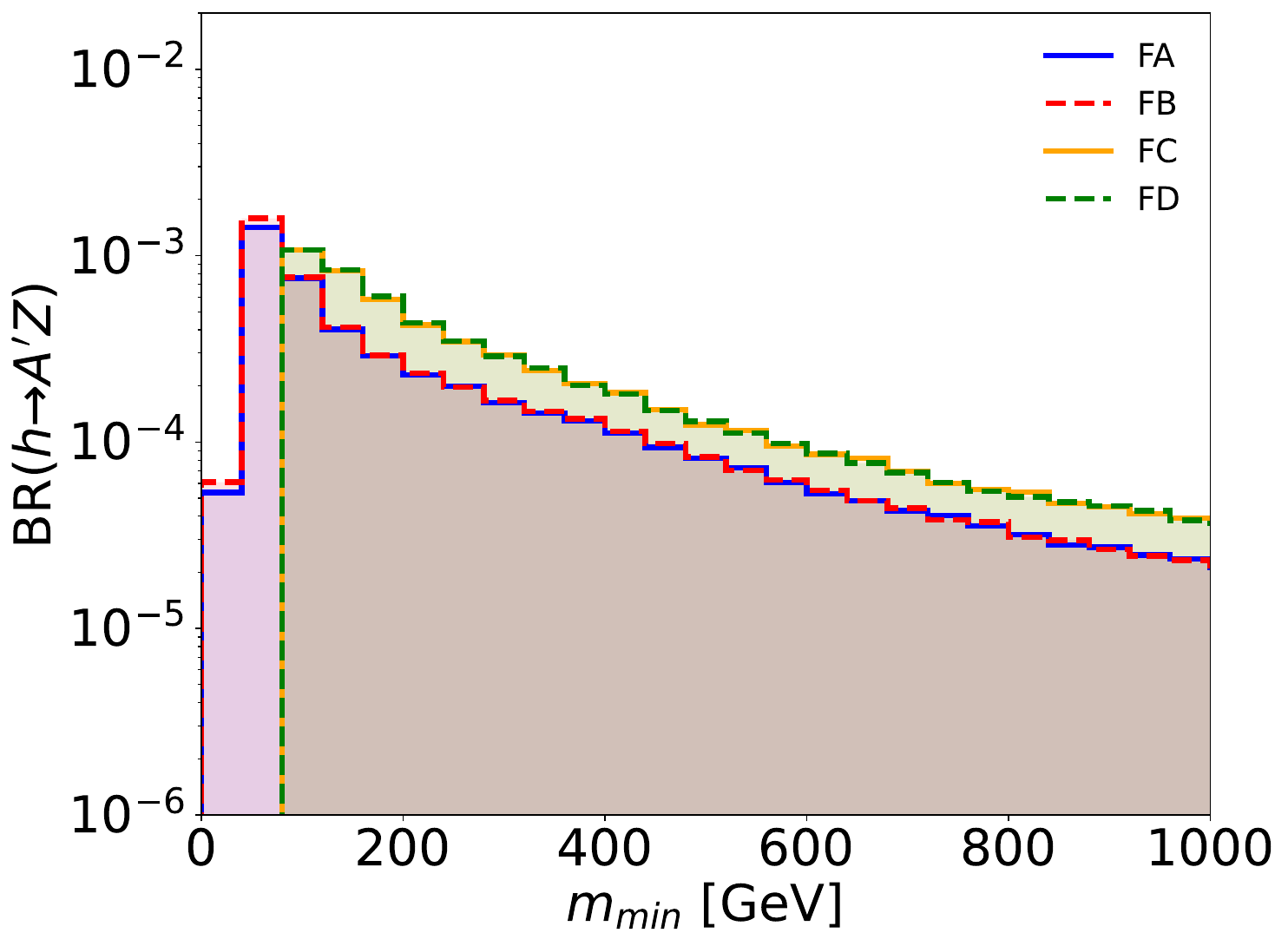}
    \label{fig:FBRApZ}
 \end{subfigure}
 \captionsetup[subfigure]{justification=centerlast}
 \begin{subfigure}{0.49\textwidth}
    \centering
    \caption{Scalar case I}
    \includegraphics[width=1\textwidth]{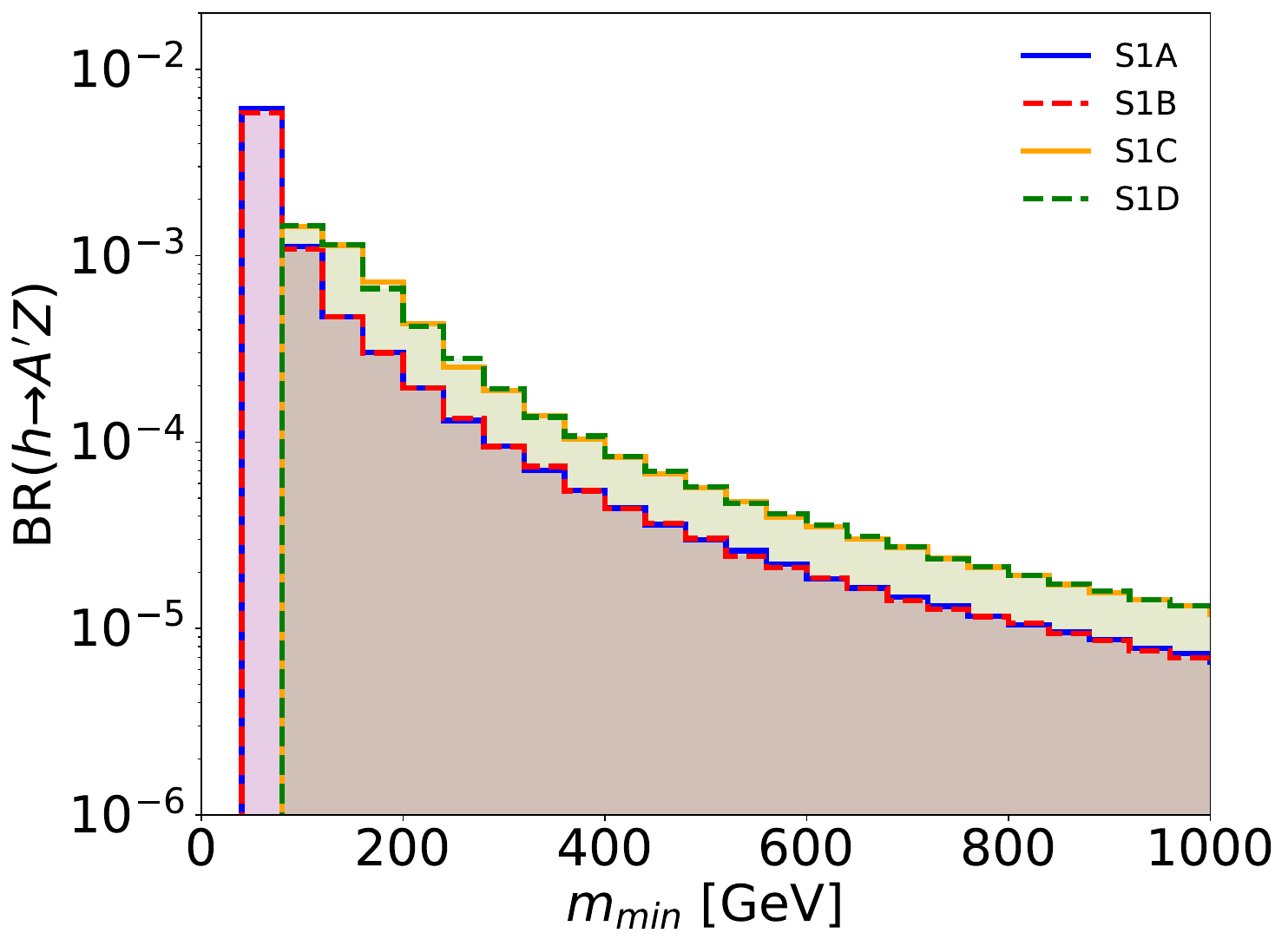}
    \label{fig:S1BRApZ}
 \end{subfigure}
 \captionsetup[subfigure]{justification=centerlast}
 \begin{subfigure}{0.49\textwidth}
    \centering
    \caption{Scalar case II}
    \includegraphics[width=1\textwidth]{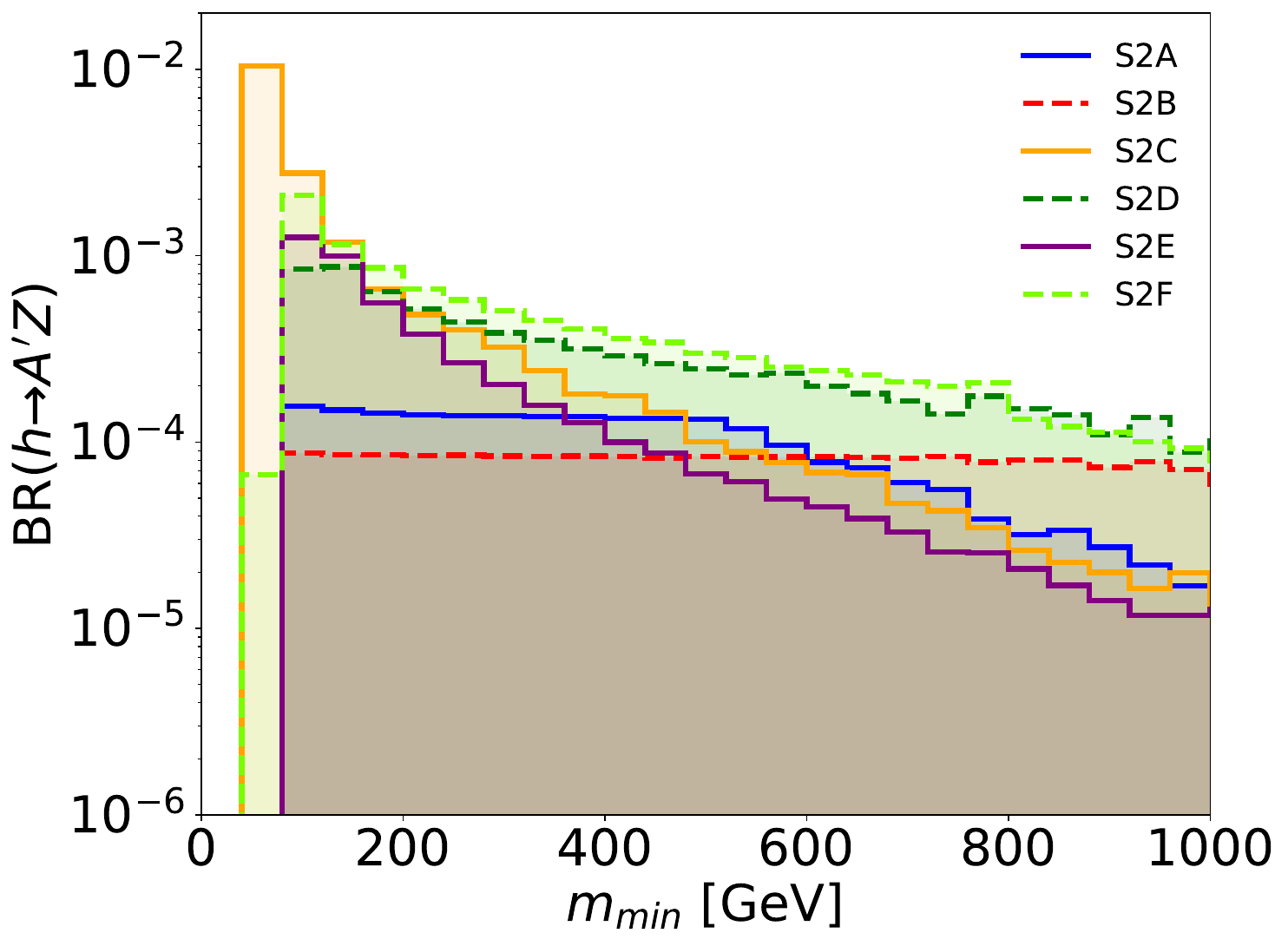}
    \label{fig:S2BRApZ}
 \end{subfigure}
 \captionsetup[subfigure]{justification=centerlast}
 \begin{subfigure}{0.49\textwidth}
    \centering
    \caption{Scalar case III}
    \includegraphics[width=1\textwidth]{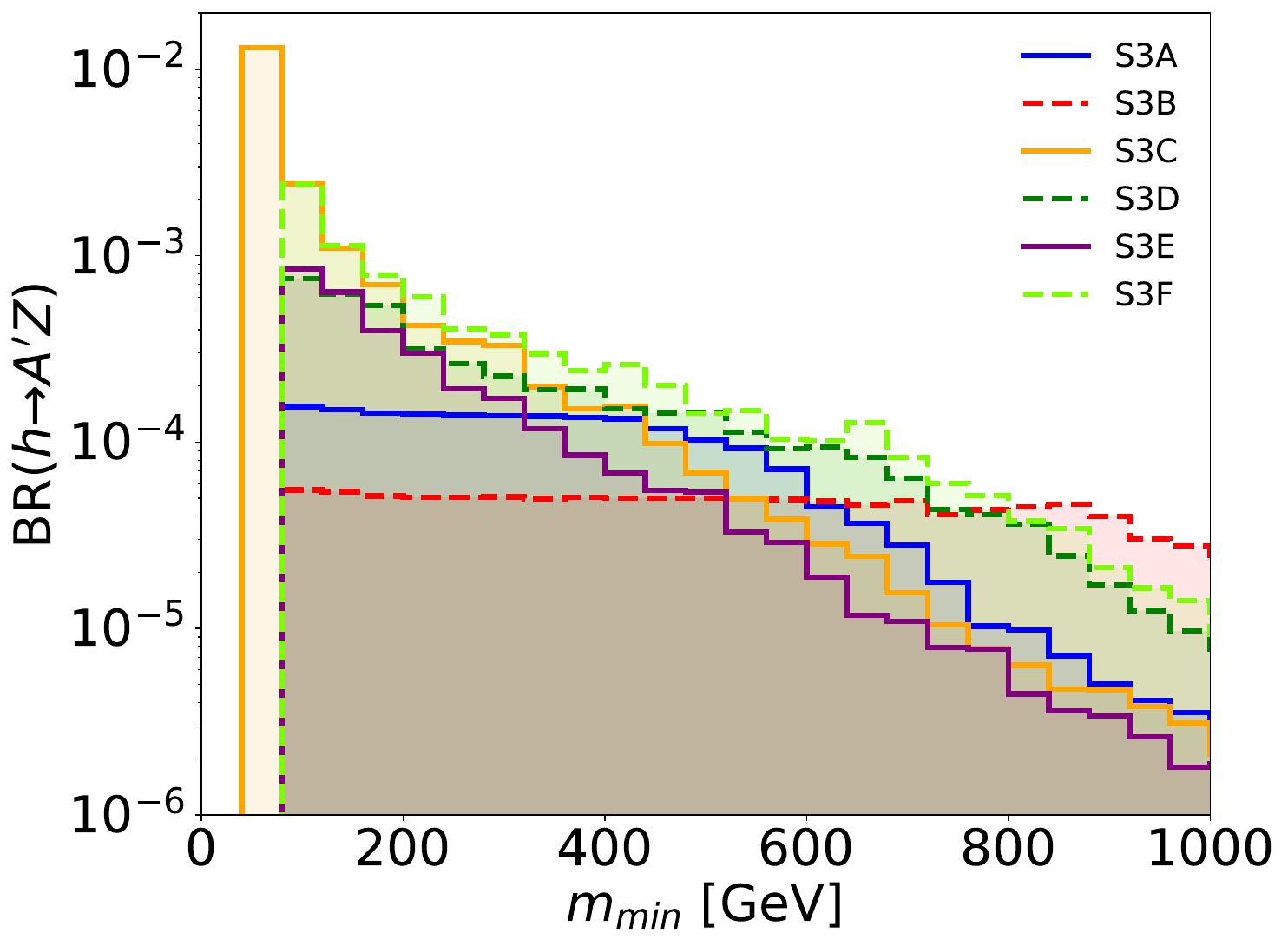}
    \label{fig:S3BRApZ}
 \end{subfigure}
  \captionsetup[subfigure]{justification=centerlast}
 \begin{subfigure}{0.49\textwidth}
    \centering
    \caption{Scalar case IV}
    \includegraphics[width=1\textwidth]{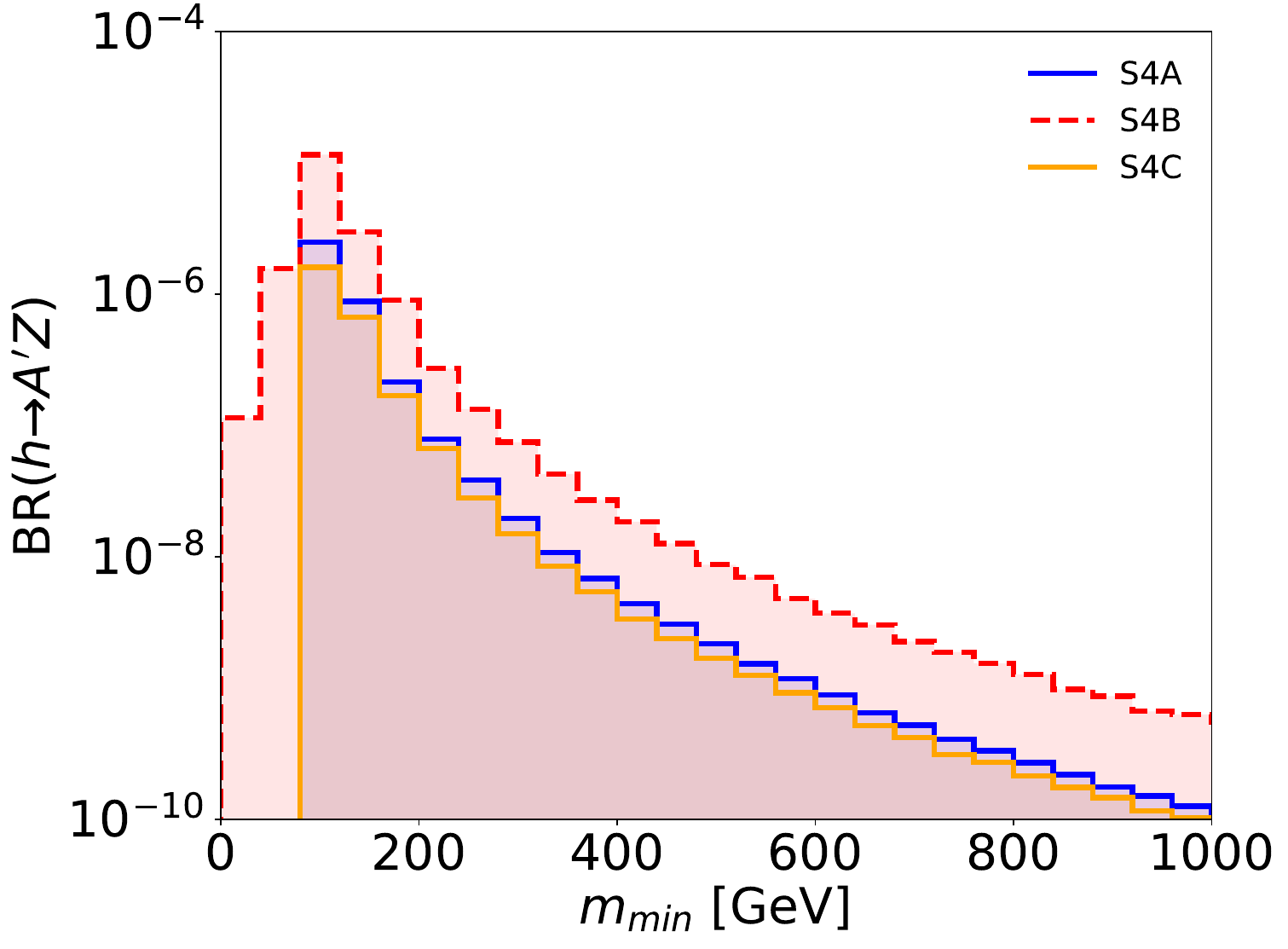}
    \label{fig:S4BRApZ}
 \end{subfigure}
  \captionsetup[subfigure]{justification=centerlast}
\caption{Allowed range of $\text{BR}(h \to A'Z)$ for different cases of a single interaction term. See Tables~\ref{table:BenchmarksI} and~\ref{table:BenchmarksII} for a description of the benchmarks. The missing benchmarks do not contribute to this decay.}
\label{fig:BRApZ}
\end{center}
\end{minipage}
\end{figure*}

Second, Fig.~\ref{fig:BRApZ} displays the allowed range of $\text{BR}(h \to A' Z)$. As can be seen, this branching ratio could in principle reach values above 1$\%$. However, doing so would require new neutral particles just above half the mass of the Higgs boson, which requires careful fine-tuning between the different parameters and is not possible for every model. Otherwise, a branching ratio of $\mathcal{O}(0.1\%)$ is relatively easy to obtain. We mention that the coefficient $S^{h \to A' Z}$ is less constrained than $S^{h \to A' A}$. As alluded to before, the reason that the constraints on $\text{BR}(h \to A' Z)$ are generally stronger than those on $\text{BR}(h \to A' A)$ of Refs.~\cite{Beauchesne:2022fet, Beauchesne:2023bcy} is the phase-space suppression in Eq.~\eqref{eq:hAZwidth}. The branching ratio $S^{h \to A' Z}$ is generally optimized for couplings of the dark photon $A'_D$ or $A'_C$ of $\mathcal{O}(1)$, with the exact value being model dependent.

\begin{figure}[t!]
\begin{center}
  \includegraphics[width=0.6\textwidth]{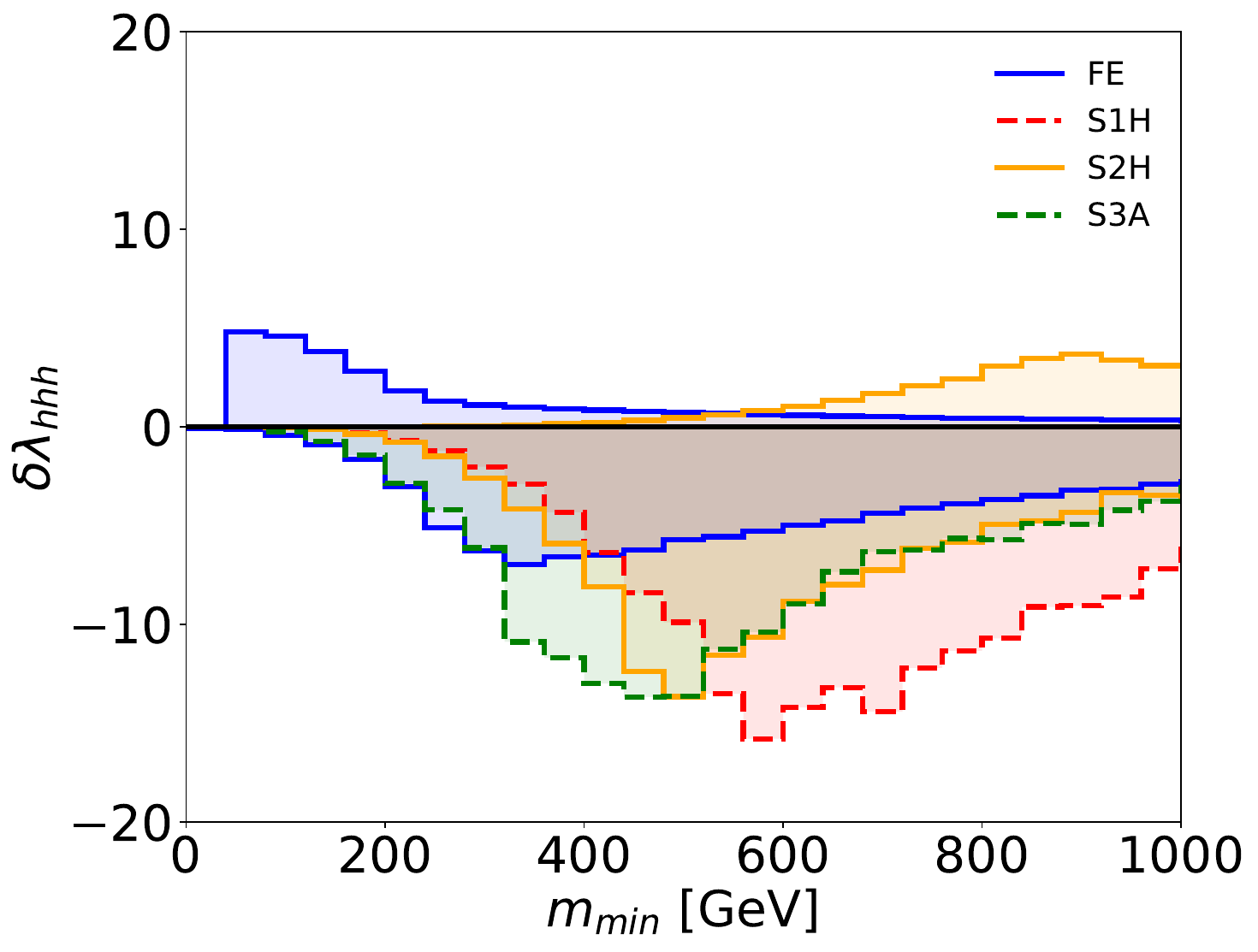}
  \caption{Examples of allowed $\delta\lambda_{hhh}$ range. See Tables~\ref{table:BenchmarksI} and~\ref{table:BenchmarksII} for a description of the benchmarks.}
\label{fig:deltahhh}
\end{center}
\end{figure}

Third, Fig.~\ref{fig:deltahhh} shows examples of the relative deviation of the triple Higgs coupling $\delta\lambda_{hhh}$ as a function of $m_{\text{min}}$. As can be seen, certain inert multiplets can easily give large contributions to the triple Higgs couplings, including both positive and negative ones. Even larger corrections are possible, though current constraints on the Higgs self-interactions render such numbers less relevant \cite{ATLAS:2022jtk}.

We mention that constraints on direct production of the new particles would generally apply to a fully defined model. However, doing so in our case would require both specifying the Lagrangian beyond the interaction terms and searches that do not presently exist, making it practically unfeasible. If this were possible, constraints could be enhanced.

\subsection{Multiple interaction terms}\label{sSec:MultipleVerticesResults}
\begin{figure*}[t!]
\begin{minipage}[b]{.97\textwidth}
\begin{center}
 \captionsetup[subfigure]{justification=centerlast}
 \begin{subfigure}{0.49\textwidth}
    \centering
    \caption{$\delta\text{BR}(h \to A Z)$}
    \includegraphics[width=1\textwidth]{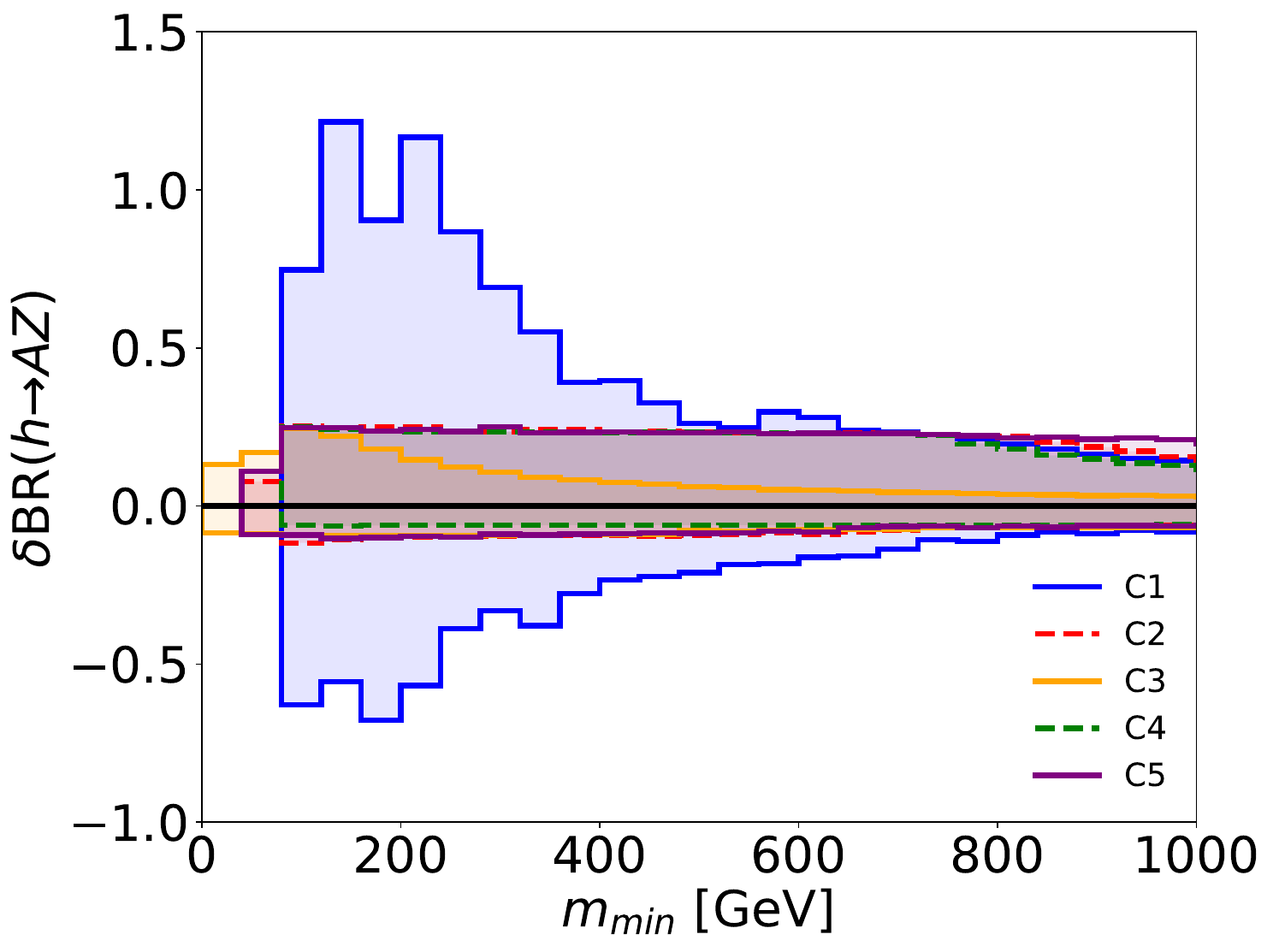}
    \label{fig:CBRAZ}
 \end{subfigure}
 \captionsetup[subfigure]{justification=centerlast}
 \begin{subfigure}{0.49\textwidth}
    \centering
    \caption{$\text{BR}(h \to A' Z)$}
    \includegraphics[width=1\textwidth]{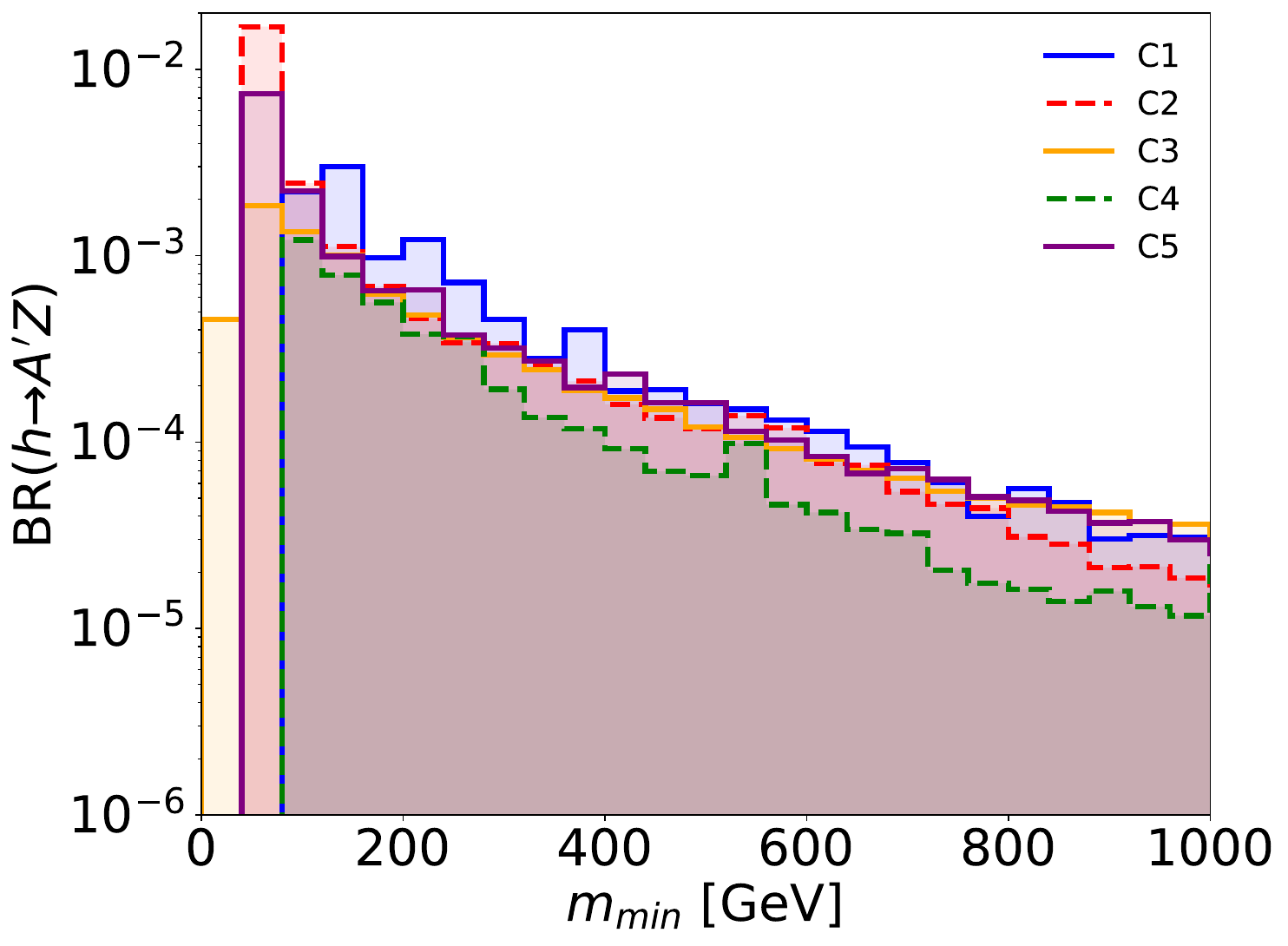}
    \label{fig:CBRApZ}
 \end{subfigure}
\caption{Allowed ranges of (a) $\delta\text{BR}(h \to A Z)$ and (b) $\text{BR}(h \to A' Z)$ for the benchmarks with multiple interactions. See Table~\ref{table:BenchmarksIII} for a description of the benchmarks.}
\label{fig:MultiVertex}
\end{center}
\end{minipage}
\end{figure*}

We now present results in the presence of multiple interactions of the types presented in Sec.~\ref{Sec:Vertices}. The chosen benchmarks are presented in Table~\ref{table:BenchmarksIII} and the results in Fig.~\ref{fig:MultiVertex}.

\begin{table}[t!]
\centering
\begin{tabular}{m{1cm} m{1cm} m{2.6cm} m{1.6cm} m{5cm}}
 \hline
 Name & Fields                          & Gauge numbers                                                          & Type              & Lagrangian            \\
 \hline
 C1  & $\phi_1$ \hspace{2cm} $\phi_2$  & $(\mathbf{3},   0, 1)$ \;$(\mathbf{1},    1, 1)$ & Complex   Complex  & $S_2(\phi_1) + S_2(\phi_2)$ \\ 
 \hline
 C2  & $\phi_1$ \hspace{2cm} $\phi_2$  & $(\mathbf{3},   0, 1)$\; $(\mathbf{2}, 1/2, 1)$ & Complex   Complex  & $S_2(\phi_1) + S_2(\phi_2) + S_1(\phi_2, \phi_1)$ \\ 
 \hline
 C3  & $\psi_1$ \hspace{2cm} $\psi_2$  \hspace{2cm} $\psi_3$ & $(\mathbf{2}, 1/2, 1)$ $(\mathbf{3}, 0, 1)$\; $(\mathbf{1}, 0, 1)$ & Complex   Complex  Complex & $F(\psi_1, \psi_2) + F(\psi_1, \psi_3)$ \\ 
  \hline
 C4  & $\phi_1$ \hspace{2cm} $\phi_2$  & $(\mathbf{3},   0, 1)$\; $(\mathbf{3},    0, 1)$ & Complex   Complex  & $S_2(\phi_1) + S_2(\phi_2) + S_3(\phi_1, \phi_2)$ \\ 
 \hline
 C5  & $\phi_1$ \hspace{2cm} $\phi_2$  \hspace{2cm} $\phi_3$ & $(\mathbf{3}, 0, 1)$ $(\mathbf{2}, 1/2, 1)$ $(\mathbf{2}, -1/2, 1)$ & Complex   Complex  Complex & $S_2(\phi_1) + S_2(\phi_2) + S_2(\phi_3) + S_1(\phi_2, \phi_1) + S_1(\phi_1, \phi_3)$ \\ 
  \hline
\end{tabular}
\caption{Benchmarks with multiple interaction terms}
\label{table:BenchmarksIII}
\end{table}

As can be seen, the range of allowed $\delta\text{BR}(h \to A Z)$ can be considerably extended. However, this requires careful tuning between different interaction terms to ensure that their contributions to constrained observables interfere destructively and constructively for $\delta\text{BR}(h \to A Z)$. This is why much larger values of $\delta\text{BR}(h \to A Z)$ are allowed for combinations of interactions than with a single interaction term. This is reflected in practice by the numerical difficulty in obtaining such large values and is also why the contours are less smooth in these plots. Additionally, such careful tuning is not even possible for all models, as certain combinations of interaction terms do not allow for a larger range than the individual terms do.  We do not observe any qualitative enhancement in the allowed range of $\text{BR}(h \to A' Z)$.

\section{Conclusion}\label{Sec:Conclusion}
Inert multiplets can only interact with the Higgs boson via a finite number of interaction terms and the forms of these terms are enough to determine their leading contributions to many Higgs properties. In this paper, we have studied the contributions of inert multiplets to the branching ratios of the Higgs to a $Z$ boson and either a photon or a dark photon and to the triple Higgs coupling.

We reach the following conclusions. The branching ratio $\text{BR}(h \to A Z)$ can deviate from its SM value by $\mathcal{O}(20\%)$ even for simple models of inert multiplets. Larger deviations are possible, but require complicated models and precise fine-tuning. Therefore, the current measurement of $\text{BR}(h \to A Z)$ being $2.1 \pm 0.7$ larger than the Standard Model value might be explainable by inert multiplets, but any such model would have to be contrived and very fine-tuned.

The branching ratio $\text{BR}(h \to A' Z)$ could in principle reach above $1\%$. However, this would again require considerable fine-tuning, and values of $\mathcal{O}(0.1\%)$ are more realistic. Such small branching ratios are difficult to probe at the LHC for two reasons. First, hadronic decays of the $Z$ boson would be difficult to reconstruct and its branching ratio to leptons is small. Second, the fact that the mass of the $Z$ boson is not that far off from that of the Higgs boson would lead to much less missing transverse momentum than for $h \to A A'$. It therefore seems unlikely that this decay channel would be observable at the LHC.

Large deviations on the triple Higgs coupling are however perfectly possible and could be an ideal observable to probe the inert multiplet scenarios considered in this work. Though a full study is beyond the scope of this paper, it seems to indicate that a first-order electroweak phase transition could be explained by inert multiplets.

\acknowledgments
This work was supported by the National Science and Technology Council under Grant No. NSTC-111-2112-M-002-018-MY3, the Ministry of Education (Higher Education Sprout Project NTU-112L104022), and the National Center for Theoretical Sciences of Taiwan.

\bibliography{biblio}
\bibliographystyle{utphys}

\end{document}